\documentstyle[10pt]{article}
\begin{document}

\hsize = 15.0 truecm \vsize = 24 truecm

\def\thf{\baselineskip=\normalbaselineskip\multiply\baselineskip
by 7\divide\baselineskip by 6}
\thf
\noindent
{\it Lectures in FORMATION AND INTERACTIONS OF TOPOLOGICAL DEFECTS,\\
NATO ASI {\bf B349}, ed A.-C. Davis, R. Brandenberger, pp 303-348 \\ 
(Plenum, New York, 1995).}

\hskip 1.2 cm

\noindent
{\bf DYNAMICS OF COSMIC STRINGS AND OTHER BRANE MODELS}

\vskip 0.8 cm
\noindent
{\bf Brandon Carter. }

\vskip 0.8 cm
\noindent
C.N.R.S., D\'epartment d'Astrophysique Relativiste et de
Cosmologie, \\
Observatoire de Paris, 92195 Meudon, FRANCE.
\vskip 1.4 cm

{\bf Abstract.}The supporting worldsheet of a string, membrane, or other 
higher dimensional brane, is analysed in terms of its first, second, and
third fundamental tensors, and its inner and outer curvature tensors. The
dynamical equations governing the models appropriate for phenomena such as
(superconducting) cosmic strings and cosmic domain walls are developed in a
general framework (allowing for both electromagnetic and Kalb Ramond
background coupling). It is shown how the surface stress momentum energy
density tensor determines the propagation characteristics of small ``wiggle"
perturbations of the worldsheet. Attention is then focussed on special
features of strings  (using the transonic model with tension $T$ inversely
proportional to the energy density  $U$ as a particularly important
example). A quadratic Hamilton-Jacobi formulation is shown to govern
equilibium states and other conservative string configurations sharing a
symmetry of the (gravitational, electromagnetic, and Kalb-Ramond)
background, including stable ring states that may be cosmologically
important.   

\vskip 1.4 cm

\def\fff{\baselineskip=\normalbaselineskip}
%$ \vbox {\offinterlineskip{\hbox {$^{(p)}$} \hbox{T} }} $.

\def\spose#1{\hbox to 0pt{#1\hss}} 
\def\Libra{\spose {--} {\cal L}}
\def\Diam{\spose {\raise 0.3pt\hbox{+}} {\diamondsuit}  }

\def\eqdef{\fff\ \vbox{\hbox{$_{_{\rm def}}$} \hbox{$=$} }\ \thf }
\def\ov{\overline}
\def\oov{\offinterlineskip\overline}
\def\oc{\check}
\def\ot{\widetilde}
\def\dL{d_{_{\rm L}}}

\def\uth{{^{\,_{(3)}}}\!}  \def\utw{{^{\,_{(2)}}}\!}
\def\ud{{^{\,_{(\rm d)}}}\!}  \def\udi{{^{\,_{(\rm d-1)}}}\!}
\def\up{{^{\,_{(\rm p)}}}\!}  \def\udp{{^{\,_{(\rm d)}}}\!}

\def\og{\eta}

\def\nabl{\nabla\!} \def\onab{\ov\nabl}

\def\vv{_{\ \vert}}     \def\osb{\offinterlineskip\hbox}
\def\af{\fff\vbox{\osb{$_{_{\vv}}$} \hbox{$\ov {\ f} $} }\thf }
\def\ag{\perp\!}

\def\P{{\cal P}} \def\R{{\cal R}} \def\W{{\cal W}} \def\E{{\cal E}} 
\def\I{{\cal I}} \def\L{{\cal L}}
\def\A{{_A}} \def\B{{_B}} \def\X{{_X}} \def\Y{{_Y}}

\def\lag{{\cal L}}
\def\aag{{m^{\rm d}}} \def\bag{{\rm b}}   \def\cag{{\rm c}}

\def\pag{{\cal P}} \def\kag{{\mit\Sigma}}

\section{Worldsheet Curvature Analysis}

In preparation for the more specific study of strings in the last three
sections of this course, the first three sections are intended as an
introduction to the systematic study, in a classical relativistic framework,
of ``branes", meaning physical models in which the relevant fields are
confined to supporting worldsheets of lower dimension than the background
spacetime. While not entirely new\cite{[1]}\cite{[2]}, this subject is still
at a rather early stage of development (compared with the corresponding 
quantum theory\cite{[3]} which has been stimulated by the  rise of
 ``superstring theory"), the main motivation for recent work\cite{[4]} on 
classical relativistic brane theory being its application to vacuum defects 
produced by the Kibble mechanism\cite{[5]}, particularly when of composite type as 
in the case of cosmic strings attached to external domain walls\cite{[6]}
and of cosmic strings carrying internal currents\cite{[7]}.

Before discussing the dynamic laws governing the evolution of a brane
worldsheet it is worthwhile to devote this first section to a recapitulation
of the essential differential geometric machinery\cite{[8]}\cite{[9]} needed
for the analysis of a timelike worldsheet of dimension d say in a background
space time manifold of dimension n. At this stage no restriction will be 
imposed on the curvature of the metric -- which will as usual be represented
with respect to local background coordinates $x^\mu$ ($\mu$= 0, ..., n--1)
by its components $g_{\mu\nu}$ -- though it will be postulated to be flat,
or at least stationary or conformally flat, in many of the applications to be 
discussed later.

The development of geometrical intuition and of computationally efficient
methods for use in string and membrane theory has been hampered by a
tradition of publishing results in untidy, highly gauge dependent, notation
(one of the causes being the undue influence still exercised by Eisenhart's
obsolete treatise ``Riemannian Geometry"\cite{[10]}). For the intermediate steps in
particular calculations it is of course frequently useful and often
indispensible to introduce specifically adapted auxiliary structures, such
as curvilinear worldsheet coordinates $\sigma^i$ ($i$= 0, ... , d--1)
and the associated bitensorial derivatives
$$x^\mu_{\ ,i}={\partial x^\mu\over\partial\sigma^i}\ ,\eqno(1.1)$$
or specially adapted orthonormal frame vectors, consisting of an internal 
subset of vectors $\iota_{\!\A}{^\mu}$ ({ A}= 0,  ... , d--1) 
tangential to the worldsheet and an external subset of vectors 
$\lambda_{\X}{^\mu}$ ({ X} = 1, ... , n--d) orthogonal to the 
worldsheet, as characterised by
$$\iota_{\!\A}{^\mu}\iota_{\!\B\mu}=\eta_{\A\B} \ ,\hskip 1 cm 
\iota_{\!\A}{^\mu}\lambda_{\X\mu}=0\ ,\hskip 1 cm 
\lambda_{\X}{^\mu}\lambda_{\Y\mu}=\delta_{\X\Y}\ ,\eqno(1.2)$$
where $\eta_{\A\B}$ is a fixed d-dimensional Minkowski metric and the
Kronecker matrix $\delta_{\X\Y}$ is a fixed (n--d)-dimensional Cartesion
metric. Even in the most recent literature there are still (under
Eisenhart's uninspiring influence) many examples of insufficient effort to 
sort out the messy clutter of indices of different kinds (Greek or Latin,
early or late, small or capital) that arise in this way by grouping the
various contributions into simple tensorially covariant combinations. Another
inconvenient feature of many publications is that results have been left 
in a form that depends on some particular gauge choice (such as the 
conformal gauge for internal string coordinates) which obscures the 
relationship with other results concerning the same system but in a 
different gauge.

The strategy adopted here\cite{[119]} aims at minimising such problems
(they can never be entirely eliminated) by working as far as possible with a
single kind of tensor index, which must of course be the one that is most
fundamental, namely that of the background coordinates, $x^\mu$. Thus, to
avoid dependence on the internal frame index {A} (which is
lowered and raised by contraction with the fixed d-dimensional Minkowski
metric $\eta_{\A\B}$ and its inverse $\eta^{\A\B}$) and on the external
frame index {X} (which is lowered and raised by contraction with
the fixed (n-d)-dimensional Cartesian metric $\delta_{\X\Y}$ and its inverse
$\delta^{\X\Y}$), the separate internal frame vectors $\iota_{\!\A}{^\mu}$
and external frame vectors $\lambda_{\X}{^\mu}$ will as far as possible be
eliminated in favour of the frame gauge independent combinations
$$\og^\mu_{\ \nu}=\iota_{\!\A}{^\mu}\iota^{\A}{_\nu} \ ,\hskip 1 cm
\ag^{\!\mu}_{\,\nu}=\lambda_{\X}{^\mu}\lambda^{\!\X}{_\nu}\ , \eqno(1.3)$$
of which the former, $\og^\mu_{\ \nu}$, is what will be referred to as the
(first) {\it fundamental tensor} of the metric, which acts as the (rank d)
operator of tangential projection onto the world sheet, while the latter,
$\ag^{\!\mu}_{\,\nu}$, is the complementary  (rank n--d) operator of 
projection orthogonal to the world sheet. 

The same principle applies to the avoidance of unnecessary involvement of
the internal coordinate indices which are lowered and raised by contraction
with the induced metric on the worldsheet as given by
$$ h_{ij}=g_{\mu\nu} x^\mu_{\ ,i} x^\nu_{\ ,j}\ , \eqno(1.4)$$
and with its contravariant inverse $h^{ij}$.
After being cast (by index raising if necessary) into its contravariant form,
any internal coordinate tensor can be directly projected onto a corresponding
background tensor in the manner exemplified by the intrinsic metric
itself, which gives 
$$\og^{\mu\nu}= h^{ij} x^\mu_{\ ,i} x^\nu_{\ ,j} \ ,\eqno(1.5)$$
thus providing an alternative (more direct) prescription for the fundamental
tensor that was previously introduced via the use of the internal frame in 
(1.3). This approach also provides a direct prescription for the orthogonal
projector that was introduced via the use of an external frame in
(1.3) but that is also obtainable immediately from (1.5) as
$$\ag^{\!\mu}_{\,\nu}=g^\mu_{\ \nu}-\og^\mu_{\ \nu}\ .\eqno(1.6)$$

In so far as we are concerned with tensor fields such as the frame vectors
whose support is confined to the d-dimensional world sheet, the effect of
Riemannian covariant differentation $\nabl_\mu$ along an arbitrary
directions on the background spacetime  will not be well defined, only the
corresponding tangentially projected differentiation operation
$$ \onab_\mu\eqdef\og\,^\nu_{\ \mu}\nabl_\nu\ , \eqno(1.7)$$
being meaningful for them, as for instance in the case of a scalar field
$\varphi$ for which the tangentially projected gradient is given in
terms of internal coordinate differentiation simply by
$\ov\nabla{^\mu}\varphi=h^{ij} x^\mu{_{\!,i}}\,\varphi_{,j}\,$.

An irreducible basis for the various possible covariant derivatives of 
the frame vectors consists of the {\it internal rotation} pseudo-tensor
$\rho_{\mu\ \rho}^{\,\ \nu}$ and the {\it external rotation} (or ``twist")
pseudo-tensor $\omega_{\mu\ \rho}^{\,\ \nu}$ as given by
$$\rho_{\mu\ \rho}^{\,\ \nu}=\og^\nu\!{_\sigma}\,\iota^{\A}{_\rho}
\onab_\mu\, \iota_{\!\A}{^\sigma}=-\rho_{\mu\rho}{^\nu}\ ,\hskip 1 cm
\omega_{\mu\ \rho}^{\,\ \nu}=\ag^{\!\nu}_{\,\sigma}\,\lambda^{\!\X}{_\rho}
\onab_\mu \lambda_{\X}{^\sigma}=-\omega_{\mu\rho}{^\nu}\ , \eqno(1.8)$$
together with the {\it second fundamental tensor} $K_{\mu\nu}{^\rho}$
as given by
$$K_{\mu\nu}{^\rho}=\ag^{\!\rho}_{\,\sigma}\,\iota^{\A}{_\nu}
\onab_\mu\, \iota_{\!\A}{^\sigma}=-\og^\sigma\!{_\nu}\,\lambda_{\X}{^\rho}
\onab_\mu \lambda^{\!\X}{_\sigma}\ .\eqno(1.9)$$
The reason for qualifying the fields (1.8) as ``pseudo" tensors is that
although they are tensorial in the ordinary sense with respect to changes
of the background coordinates $x^\mu$ they are not geometrically well 
defined just by the geometry of the world sheet but are gauge dependent
in the sense of being functions of the choice of the internal and external
frames $\iota_{\!\A}{^\mu}$ and $\lambda_{\X}{^\mu}$. On the other hand,
like the first fundamental tensor $\og^{\mu\nu}$ as given by (1.5), the
second fundamental tensor (1.9) is geometrically well defined
in the sense of being frame gauge independent, as can be seen from
the  equivalent but more direct definition\cite{[4]}
$$K_{\mu\nu}{^\rho} \eqdef \og\,^\sigma_{\ \nu}\onab_\mu
\og\,^\rho_{\ \sigma} \ .\eqno(1.10)$$

The gauge dependence of the rotation tensors $\rho_{\mu\ \rho}^{\,\ \nu}$
and $\omega_{\mu\ \rho}^{\,\ \nu}$ means that (unlike $K_{\mu\nu}{^\rho}$)
they can each be set to zero at any particular given point on the worldsheet
by choice of the relevant frames in its vicinity. However the condition for
it to be possible to set these pseudo-tensors to zero throughout an open
neigbourhood is the vanishing of the curvatures of the corresponding frame
bundles as characterised with respect to the respective invariance subgroups
SO(1,d--1) and SO(n--d) into which the full Lorentz invariance group 
SO(1,n--1)
is broken by the specification of the d-dimensional world sheet orientation.
The {\it inner curvature} that needs to vanish for it to be possible for
$\rho_{\mu\ \rho}^{\,\ \nu}$ to be set to zero in an open neighbourhood is
of Riemannian type, is obtainable (by a calculation of the type originally
developed by Cartan that was made familiar to physicists by Yang Mills
theory) as\cite{[8]}
$$R_{\kappa\lambda}{^\mu}{_\nu}=2\og^\mu{_{\!\sigma}}\og^\tau{_{\!\mu}}
\og^\pi{_{[\lambda}}\onab_{\kappa]}\rho_{\pi\ \tau}^{\,\ \sigma}+2
\rho_{[\kappa}{^{\mu\pi}}\rho_{\lambda]\pi\nu}\ ,\eqno(1.11)$$
while the {\it outer curvature} that needs to vanish for it to be possible
for the ``twist" tensor $\omega_{\mu\ \rho}^{\,\ \nu}$ to be set to zero in 
an open neighbourhood is of a less familiar type that is given\cite{[8]} by
$$\Omega_{\kappa\lambda}{^\mu}{_\nu}=2\ag^{\!\mu}_{\,\sigma}\ag^{\!\tau}
_{\,\mu}\,\og^\pi{_{[\lambda}}\onab_{\kappa]}\omega_{\pi\ \tau}^{\,\ \sigma}
+2\omega_{[\kappa}{^{\mu\pi}}\omega_{\lambda]\pi\nu}\ .\eqno(1.12)$$
The frame gauge invariance of the expressions (1.11) and (1.12) is not
immediately obvious, but will be made manifest in the the alternative
expressions to be given below following a synopsis of the properties of 
the (by (1.10) manifestly) gauge invariant second fundamental tensor 
$K_{\mu\nu}{^\rho}$.

An equation of the form (1.10) for $K_{\mu\nu}{^\rho}$ is of course meaningful
not only for the fundamental projection tensor of a d-surface, but also
for any (smooth) field of rank-d projection operators $\og\,^\mu_{\ \nu}$
as specified by a field of arbitrarily orientated d-surface elements. What
distinguishes the integrable case, i.e. that in which the elements mesh
together to form a well defined d-surface through the point under
consideration, is the condition that the tensor defined by (1.10) should also
satisfy the {\it Weingarten identity}
$$K_{[\mu\nu]}{^\rho} =0 \eqno(1.13)$$ 
(where the square brackets denote antisymmetrisation), this symmetry
property of the second fundamental tensor being derivable\cite{[4]}\cite{[8]}
as a version of the well known Frobenius theorem. In addition to this 
non-trivial symmetry property, the second fundamental tensor is also obviously
tangential on the first two indices and almost as obviously orthogonal on
the last, i.e. 
$$\ag^{\!\sigma}_{\,\mu}K_{\sigma\nu}{^\rho}=K_{\mu\nu}{^\sigma}
\og_\sigma{^\rho}=0   \ . \eqno(1.14)$$ 
The second fundamental tensor $K_{\mu\nu}{^\rho}$ has the property of fully
determining the tangential derivatives of the first fundamental tensor
$\og\,^\mu_{\ \nu}$ by the formula
$$\onab_\mu\og{_{\nu\rho}}=2K_{\mu(\nu\rho)} \eqno(1.15)$$ 
(using round brackets to denote symmetrisation) and it can be seen to be
characterisable by the condition that the orthogonal projection of the
acceleration of any tangential vector field $u^\mu$ will be given by
$$u^\mu u^\nu K_{\mu\nu}{^\rho}=\ag^{\!\rho}_{\,\mu}\dot u^\mu\ ,
\hskip 1 cm \dot u^\mu={u^\nu\nabla{_\nu} u^\mu} \ . \eqno(1.16)$$

It is very practical for a great many purposes to introduce the {\it
extrinsic curvature vector} $K^\mu$, defined as the trace of the second
fundamental tensor, which is automatically orthogonal to the worldsheet, 
$$ K^\mu\eqdef K^\nu_{\ \nu}{^\mu}
\ , \hskip 1 cm  \og^\mu_{\ \nu}K {^\nu}=0 \ .\eqno(1.17)$$ 
It is useful for many specific purposes to work this out in terms of the
intrinsic metric $h_{ij}$ and its determinant $\vert h\vert $. It suffices 
to use the simple expression 
$\onab^{\,\mu}\varphi=h^{ij} x^\mu{_{,i}}\varphi_{,j}$ 
for the tangentially projected gradient of a scalar field $\varphi$ on the
worldsheet, but for a tensorial field (unless one is using Minkowski
coordinates in a flat spacetime) there will also be contributions involving
the background Riemann Christoffel connection 
$$\Gamma_{\mu\ \rho}^{\,\ \nu}=g^{\nu\sigma}\big(g_{\sigma(\mu,\rho)}
-{_1\over^2}g_{\mu\rho,\sigma}\big)\ .\eqno(1.18)$$
The curvature vector is thus obtained in explicit detail as
$$K^\nu=\onab_\mu\og^{\mu\nu}={1\over\sqrt{\Vert h\Vert}}\Big(
\sqrt{\Vert h\Vert}h^{ij}x^\nu_{\, ,i}\Big){_{,j}}+h^{ij}x^\mu_{\, ,i}
x^\rho_{\, ,j}\Gamma_{\mu\ \rho}^{\,\ \nu}\ . \eqno(1.19)$$
This last expression is  technically useful for certain specific
computational purposes, but it must be remarked that much of the 
literature on cosmic string dynamics has been made unnecessarily heavy 
to read by a tradition of working all the time with long strings of non
tensorial terms such as those on the right of (1.19) rather than taking
advantage of such more succinct tensorial expressions as the preceeding 
formula $\onab_\mu\og^{\mu\nu}$. As an alternative to the universally 
applicable tensorial approach advocated here, there is of course another more 
commonly used method of achieving succinctness in particular circumstances,
which is to sacrifice gauge covariance by using specialised kinds of 
coordinate system. In particular for the case of a string, i.e. for a 
2-dimensional worldsheet, it is standard practise to use conformal 
coordinates $\sigma^{_0}$ and $\sigma^{_1}$ so that the corresponding 
tangent vectors $\dot x^\mu=x^\mu_{\, ,_0}$ and $x^{\prime\mu}= 
x^\mu_{\, ,_1}$ satisfy the restrictions $\dot x^\mu x^\prime_{\, \mu}=0$, 
$\dot x^\mu\dot x_\mu+x^{\prime\mu}x^\prime_{\,\mu}=0$, which implies 
$\sqrt{\Vert h\Vert}=x^{\prime\mu}x^\prime_{\,\mu}=-\dot x^\mu\dot x_\mu$ 
so that (1.19) simply gives $\sqrt{\Vert h \Vert}\,K^\nu=$ 
$x^{\prime\prime\nu}-\ddot x^\nu + (x^{\prime\mu}x^{\prime\rho}
-\dot x^\mu\dot x^\rho)\Gamma_{\mu\ \rho}^{\,\ \nu}$.

The physical specification of the extrinsic curvature vector (1.17) for a
timelike d-surface in a dynamic theory provides what can be taken as the
equations of extrinsic motion of the d-surface${[4][9]}$, the simplest
possibility being the ``harmonic" condition $K^\mu=0$ that is obtained (as
will be shown in the following sections) from a surface measure variational
principle such as that of the Dirac membrane model\cite{[1]}, or of the 
Goto-Nambu string model\cite{[5]} whose dynamic equations in a flat 
background are therefore expressible with respect to a standard conformal 
gauge in the familiar form $x^{\prime\prime\mu}-\ddot x^\mu=0$. There is a
certain analogy between the Einstein vacuum equations, which impose the 
vanishing of the trace $\R_{\mu\nu}$ of the background spacetime curvature 
$\R_{\lambda\mu}{^\rho}{_\nu}$, and the Dirac-Gotu-Nambu equations, which 
impose the vanishing of the trace $K^\nu$ of the second fundamental tensor 
$K_{\lambda\mu}{^\nu}$. Just as it is useful to separate out the Weyl 
tensor\cite{[11]}, i.e. the trace free part of the Ricci background curvature
which is the only part that remains when the Einstein vacuum equations are 
satisfied, so also analogously, it is useful to separate out the the trace 
free part of the second fundamental tensor, namely the extrinsic conformation
tensor\cite{[8]}, which is the only part that remains when equations of 
motion of the Dirac-Gotu-Nambu type are satisfied. Explicitly, the trace
free {\it extrinsic conformation} tensor $C_{\mu\nu}{^\rho}$ of a
d-dimensional imbedding is defined\cite{[8]} in terms of the corresponding
first and second fundamental tensors $\eta_{\mu\nu}$ and $K_{\mu\nu}{^\rho}$
as
$$C_{\mu\nu}{^\rho}\eqdef K_{\mu\nu}{^\rho}-{1\over{\rm d}}\og{_{\mu\nu}}
K^\rho \ , \hskip 1 cm     C^\nu_{\ \nu}{^\mu}=0 \ .\eqno(1.20)$$ 
Like the Weyl tensor $\W_{\lambda\mu}{^\rho}{_\nu}$ of the background metric
(whose  definition is given implicitly by (1.25) below) this conformation
tensor has the noteworthy property of being invariant with respect to
conformal modifications of the background metric:
$$g_{\mu\nu}\mapsto {\rm e}^{2\alpha}g_{\mu\nu}\ ,\hskip 0.6 cm
\Rightarrow\hskip 0.6 cm K_{\mu\nu}{^\rho}\mapsto K_{\mu\nu}{^\rho}
+\eta_{\mu\nu}\perp^{\!\rho\sigma}\nabl_\sigma\alpha\ ,\hskip 0.6 cm 
 C_{\mu\nu}{^\rho}\mapsto C_{\mu\nu}{^\rho}\ .\eqno(1.21)$$
This formula is useful\cite{[12]} for calculations of the kind undertaken
by Vilenkin\cite{[13]} in a standard Robertson-Walker type 
cosmological background, which can be obtained from a flat auxiliary 
spacetime metric by a  conformal transformation for which
${\rm e}^\alpha$ is a time dependent Hubble expansion factor.

As the higher order analogue of (1.10) we can go on to introduce the 
{\it third} fundamental tensor$\cite{[4]}$ as
$$\Xi_{\lambda\mu\nu}{^\rho} \eqdef\og\,^\sigma_{\ \mu}\og\,^\tau_{\ \nu}
\ag^{\!\rho}_{\,\alpha}\onab_\lambda K_{\sigma\tau}{^\alpha} \ , 
\eqno(1.22)$$
which  by construction is obviously symmetric between the second and third
indices and tangential on all  the first three indices.  In a spacetime 
background that is flat (or of constant curvature as is the case for the 
DeSitter universe model) this third fundamental tensor is fully symmetric 
over all the first three indices by what is interpretable as the {\it 
generalised Codazzi identity} which is expressible\cite{[8]} in a 
background with arbitrary Riemann curvature $\R_{\lambda\mu}{^\rho}{_\sigma}$ 
as 
$$\Xi_{\lambda\mu\nu}{^\rho}= \Xi_{(\lambda\mu\nu)}{^\rho} +{_2\over^3}
\og\,^\sigma_{\ \lambda}\og\,^\tau_{\ {(\mu}}  \og\,^\alpha_{\ {\nu)}}
\R_{\sigma\tau}{^\beta}{_\alpha}\ag^{\!\rho}_{\,\beta}
\  \eqno(1.23)$$
It is to be noted that a script symbol $\R$ is used here in order to
distinguish the (n-dimensional) background Riemann curvature tensor from the
intrinsic curvature tensor (1.11) of the (d-dimensional) worldship to which
the ordinary symbol $R$ has already allocated.

For many of the applications that will follow it will be sufficient just to
treat the background spacetime as flat, i.e. to take
$\R_{\sigma\tau}{^\beta}{_\alpha}=0$. At this stage however, we shall allow
for an unrestricted background curvature. For n$>2$ this will be
decomposible in terms of its trace free  Weyl part
$\W_{\mu\nu}{^\rho}{_\sigma}$ 5which as remarked above is conformally
invariant) and the corresponding background Ricci tensor and its scalar
trace,

$$\R_{\mu\nu}= \R_{\rho\mu}{^\rho}{_\nu}  \ , \hskip 1 cm
 \R=\R^\nu_{\ \nu}\ , \eqno(1.24)$$
in the form\cite{[11]}
$$\R_{\mu\nu}{^{\rho\sigma}}=\W_{\mu\nu}{^{\rho\sigma}} +{_4\over^{ n-2} }
g^{[\rho}_{\ [\mu}\R^{\sigma]}_{\ \nu]}-{_2\over ^{(n-1)(n-2)} } \R
g^{[\rho}_{\ [\mu}g^{\sigma]}_{\ \nu]} \ ,\eqno(1.25)$$
(in which the Weyl contribution can be non zero only for n$\geq$ 4).
In terms of the tangential projection of this background curvature, one can 
evaluate the corresponding {\it internal} curvature tensor (1.11) in the form
$$ R{_{\mu\nu}}{^\rho}{_\sigma}= 2K^\rho{_{[\mu}}{^\tau}
K_{\nu]\sigma\tau}+
\og\,^\kappa_{\ \mu} \og\,^\lambda_{\ \nu}\R_{\kappa\lambda}{^\alpha}{_\tau}
\og\,^\rho_{\ \alpha}\og\,^\tau_{\ \sigma}
  \ , \eqno(1.26)$$
which is the translation into the present scheme of what is well known in
other schemes as the {\it generalised Gauss identity}. The much less well 
known analogue for the (identically trace free and conformally invariant) 
{\it outer} curvature (1.12) (for which the most historically appropriate name
might be argued to be that of Schouten\cite{[11]}) is given\cite{[8]} in 
terms of the corresponding projection of the background Weyl tensor by the 
expression 
$$\Omega_{\mu\nu}{^\rho}{_\sigma}= 2C_{[\mu}{^{\tau\rho}}
C_{\nu]\tau\sigma}+
\og\,^\kappa_{\ \mu} \og\,^\lambda_{\ \nu}\W_{\kappa\lambda}{^\alpha}{_\tau}
\ag^{\!\rho}_{\,\alpha}\ag^{\!\tau}_{\,\sigma}
 \ . \eqno(1.27)$$
It follows from this last identity  that in a background that is flat or 
conformally flat (for which it is necessary, and for n$\geq 4$ sufficient, that
the Weyl tensor should vanish) the vanishing of the extrinsic conformation
tensor $C_{\mu\nu}{^\rho}$ will be sufficient (independently of the
behaviour of the extrinsic curvature vector $K^\mu$) for vanishing of the
outer curvature tensor $\Omega_{\mu\nu}{^\rho}{_\sigma}$, which is the
condition for it to be possible to construct fields of vectors $\lambda^\mu$
orthogonal to the surface and such as to satisfy the generalised
Fermi-Walker propagation  condition to the effect that $\ag^{\!\rho}_{\,\mu}
\onab_\nu\lambda_\rho$ should vanish. It can also be shown\cite{[8]}
(taking special trouble for the case d=3 )  that in a conformally flat
background (of arbitrary dimension n) the vanishing of the conformation
tensor $C_{\mu\nu}{^\rho}$ is always sufficient (though by no means
necessary) for conformal flatness of the induced geometry in the imbedding.

The application with which we shall mainly be concerned in the following
work will be the case d=2 of a string. An orthonormal tangent
frame will consist in this case just of a timelike unit vector, 
$\iota_0^{\ \mu}$, and a spacelike unit vector, $\iota_1^{\ \mu}$,
whose exterior product vector is the frame independent antisymmetric unit 
surface element tensor 
$$\E^{\mu\nu}=2\iota_0^{\ [\mu}\iota_1^{\ \nu]}
=2\big(\!-\!\vert h\vert\big)^{-1/2}\,x^{[\mu}_{\ \, ,_0}x^{\nu]}_{\ ,_1}
\ ,\eqno(1.28)$$
 whose tangential gradient satisfies
$$\onab_\lambda\E^{\mu\nu}=-2K_{\lambda\rho}{^{[\mu}}\E^{\nu]\rho}\ .
\eqno(1.29)$$
(This is the special d=2 case of what is, as far as I am aware, the only
wrongly printed formula in the more complete analysis\cite{[8]} on which this
section is based: the relevant general formula (B9) is valid as printed only
for odd d, but needs insertion of a missing sign adjustment factor
$(-1)^{\rm d-1}$ in order to hold for all d.) In this case the inner
rotation pseudo tensor (1.8) is determined just by a corresponding rotation
covector $\rho_\mu$ according to the specification
$$\rho_{\lambda\ \nu}^{\ \mu}={_1\over^2}\,\E^\mu_{\ \nu}\rho_\lambda\ ,\hskip 
1 cm \rho_\lambda=\rho_{\lambda\ \nu}^{\ \mu}\E^\nu_{\ \mu}  \ .\eqno(1.30)$$
This can be used to see from (1.11) that the Ricci scalar (1.24) of the
2-dimensional worldsheet will have the well known property
of being a pure surface divergence, albeit of a frame gauge dependent 
quantity:
$$R=\onab_\mu\big(\E^{\mu\nu} \rho_\nu\big) \ .\eqno(1.31)$$
In the specially important case of a string in ordinary 4-dimensional
spacetime, i.e. when we have not only d=2 but also n=4, the antisymmetric
background measure tensor $\varepsilon^{\lambda\mu\nu\rho}$ can be used to
determine a scalar (or more strictly, since its sign is orientation
dependent, a pseudo scalar) magnitude $\Omega$ for the outer curvature
tensor (1.12) (despite the fact that its traces are identically zero)
according to the specification
$$\Omega={_1\over^2}\,\Omega_{\lambda\mu\nu\rho}\,\varepsilon^
{\lambda\mu\nu\rho}\ ,\eqno(1.32)$$
Under these circumstances one can also define a ``twist" covector
$\omega_\mu$, that is the outer analogue of $\rho_\mu$, according to the
specification
$$\omega_\nu={_1\over^2}\,\omega_\nu^{\ \mu\lambda}\,
\varepsilon_{\lambda\mu\rho\sigma}\,\E^{\rho\sigma}\ .\eqno(1.33)$$
This can be used to deduce from (1.12) that the outer curvature (pseudo) scalar
$\Omega$ of a string worldsheet in 4-dimensions has a divergence property
of the same kind as that of its more widely known Ricci analogue (1.31),
the corresponding formula being given by
$$\Omega=\onab_\mu\big(\E^{\mu\nu} \omega_\nu\big) \ .\eqno(1.34)$$
It is to be remarked that for a compact spacelike 2-surface the integral of
(1.29) gives the well known Gauss Bonnet invariant, but that the timelike
string worldsheets under consideration here will not be characterised by any
such global invariant since they will not be compact (being open in the time
direction even for a loop that is closed in the spacial sense). The outer
analogue of the Gauss Bonnet invariant that arises from (1.32) for a spacelike
2-surface has been discussed by Penrose and Rindler\cite{[14]} but again there is
no corresponding global invariant in the necessarily non-compact timelike
case of a string worldsheet.

\section{Laws of Motion for a Regular Brane Complex}

The term p-brane has come into use\cite{[3]}\cite{[15]} to describe a dynamic
system localised on a timelike support surface of dimension d=p+1, 
imbedded in a spacetime background of dimension n$>$p. Thus at the
low dimensional extreme one has the example of a zero - brane, meaning what
is commonly referred to as a ``point particle", and of a 1-brane meaning
what is commonly referred to as a ``string". At the high dimensional extreme
one has the ``improper" case of an (n--1)-brane, meaning what is commonly
referred to as a ``medium" (as exemplified by a simple fluid), and of an
(n--2)-brane, meaning what is commonly referred to as a ``membrane" (from which 
the generic term ``brane" is derived). A membrane (as exemplified by a
cosmological domain wall) has the special feature of being supported by a
hypersurface, and so being able to form a boundary between separate
background space time regions; this means that a 2-brane has the status of
being a membrane in ordinary 4-dimensional spacetime (with $n=4$) but not in
a higher dimensional (e.g. Kaluza Klein type) background.

The purpose of the present section is to consider the dynamics not just of
an individual brane but of a {\it brane complex} or ``rigging model"\cite{[4]}
such as is illustrated by the nautical archetype in which the wind -- a
3-brane -- acts on a boat's sail -- a 2-brane -- that is held in place by
cords -- 1-branes -- which meet at knots, shackles and pulley blocks that
are macroscopically describable as point particles -- i.e. 0-branes. In
order for a a set of branes of diverse dimensions to qualify as
a``geometrically regular'' brane complex or ``rigging system" it is required
not only that the support surface of each (d--1)-brane should be a smoothly
imbedded d-dimensional timelike hypersurface but also that its boundary, if
any, should consist of a disjoint union of support surfaces of an attatched
subset of lower dimensional branes of the complex. (For example in order
qualify as part of a regular brane complex the edge of a boat's sail can not
be allowed to flap freely but must be attatched to a hem cord belonging to
the complex.) For the brane complex to qualify as regular in the strong
dynamic sense that will be postulated in the present work, it is also
required that a member p-brane can exert a direct force only on an an
attached (p--1)-brane on its boundary or on an attached (p+1)-brane on whose
boundary it is itself located, though it may be passively subject to forces
exerted by a higher dimensional background field. For instance the
Peccei-Quin axion model gives rise to field configurations representable as
regular complexes of domain walls attached to
strings\cite{[16]}\cite{[17]}\cite{[18]}, and a bounded (topological or
other) Higgs vortex defect terminated by a pair of pole
defects\cite{[19]}\cite{[20]}\cite{[21]}\cite{[22]} may be represented as a
regular brane complex consisting of a finite cosmic string with a pair of
point particles at its ends, in an approximation neglecting Higgs field
radiation. (However allowance for radiation would require the use of an
extended complex including the Higgs medium whose interaction with the
string -- and a fortiori with the terminating particles -- would violate the
regularity condition: the ensuing singularities in the back reaction would
need to be treated by a renormalisation procedure of a
kind\cite{[23]}\cite{[24]} that is beyond the scope of the present article,
but that is discussed by Paul Shellard in an accompanying article in this
volume and elsewhere{[18]}.)

The present section will be restricted to the case of a brane complex that
is not only regular in the sense of the preceeding paragraph but that is
also {\it pure} (or ``fine") in the sense that the lengthscales
characterising the internal structure of the (defect or other) localised
phenomenon represented by the brane models are short compared with those
characterising the macroscopic variations under consideration so that
polarisation effects play no role. For instance in the case of  a point
particle, the restriction that it should be describable as a ``pure" zero
brane simply means that it can be represented as a simple monopole without
any dipole or higher multipole effects. In the case of a cosmic string the
use of a ``pure" 1-brane description requires that the underlying vortex
defect be sufficiently thin compared not only compared with its total length
but also compared with the lengthscales characterising its curvature
and the gradients of any currents it may be carrying. The effect of the
simplest kind of curvature corrections beyond this ``pure brane" limit will
be discussed in Section 3, but in the rest of this article, as in the
present section, it will be assumed that the ratio of microscopic to
macroscopic lengthscales is sufficiently small for description in terms of
``pure" p-branes to be adequate.

The present section will not be concerned with the specific details
of particular cases but with the generally valid laws that can be derived
as Noether identities from the postulate that the model is governed by
dynamical laws derivable from a variational principle specified
in terms of an action  function ${\I}$. It is
however to be emphasised that the validity at a macroscopic level of the
laws given here is not restricted to cases represented by macroscopic models
of the strictly conservative type directly governed by a macroscopic
variational principle. The laws obtained here will also be applicable to
classical models of dissipative type (e.g. allowing for resistivity to
relative flow by internal currents) as necessary conditions for the
existence of an underlying variational description  of the microscopic
(quantum) degrees of freedom that are allowed for merely as entropy in the
macroscopically averaged classical description.

In the case of a brane complex, the total action $\I$ will be given as a sum
of distinct d-surface integrals respectively contributed by the various
(d--1)-branes of the complex, of which each is supposed to have its own
corresponding Lagrangian surface density scalar $\ud \L$ say. Each supporting
d-surface will be specified by a mapping $\sigma\mapsto x\{\sigma\}$ giving
the local background coordinates $x^\mu$ ($\mu$=0, .... , n--1) as functions
of local internal coordinates $\sigma^i\ $ ( i=0, ... , d--1). The
corresponding d-dimensional surface metric tensor $\ud h_{ij}$ that is
induced (in the manner described in the preceeding section) as the pull back
of the n-dimensional background spacetime metric $g_{\mu\nu}$, will
determine the natural surface measure, $\ud d\!\ov{\cal S}$, in terms of 
which the total action will be expressible in the form 
$$\I= \sum_{\rm d} \int\!\! \ud d\!\ov{\cal S}\, \ud \ov\L \ ,
 \hskip 1 cm \ud d\!\ov{\cal S}=\sqrt{\Vert^{_{\rm (d)}} h \Vert}\,
 d^{\rm d}\!\sigma \ . \eqno(2.1)$$
As a formal artifice whose use is an unnecessary complication in ordinary
dynamical calculations but that can be useful for  purposes such as the
calculation of radiation, the {\it confined} (d-surface supported) but
locally {\it regular} Lagrangian scalar fields $\ud \ov\L$ can be
replaced by corresponding unconfined, so no longer regular but {\it
distributional} fields $\ud \hat\L$, in order to allow the the basic
multidimensional action (2.1) to be represented as a single integral, 
$$\I =\int\!\! d\!{\cal S}\, \sum_{\rm d} \ud \hat\L \ ,
 \hskip 1 cm d\!{\cal S}=\sqrt{\Vert g \Vert}\, d^{\rm n}x \ .\eqno(2.2)$$
over the n-dimensional {\it background} spacetime. In order to do this, it
is evident that for each (d--1)-brane of the complex the required
distributional action contribution $\ud\hat\L$  must be constructed in terms
of the corresponding regular d-surface density scalar $\ud\ov\L$
according to the prescription that is expressible in standard Dirac notation
as
$$ \ud\hat\L=\Vert g \Vert^{-1/2}\int\!\! \ud d\!\ov{\cal S}\,
\ud\ov\L\, \delta^{\rm n}[x-x\{\sigma\}]\ .\eqno(2.3)$$

In the kind of model under consideration, each supporting d-surface is
supposed to be endowed with its own independent internal field variables
which are allowed to couple with each other and with their derivatives in
the corresponding d-surface Lagrangian contribution $\ud\ov\L$, and which
are also allowed to couple into the Lagrangian contribution $\udi\ov\L$ on
any of its attached boundary (d--1) surfaces, though -- in order not to
violate the strong dynamic regularity condition -- they are not allowed to
couple into contributions of dimension (d--2) or lower. As well as involving
its own d-brane surface fields and those of any (d+1) brane to whose
boundary it may belong, each contribution $\ud\ov\L$ may also depend
passively on the fields of a fixed higher dimensional background. Such
fields will of course always include the background spacetime metric
$g_{\mu\nu}$ itself. Apart from that, the most commonly relevant kind of
backround field (the only one allowed for in the previous analysis{[4]})
is a Maxwellian gauge potential $A_\mu$ whose exterior
derivative is the automatically ``closed" electromagnetic field,
$$F_{\mu\nu}=2\nabl_{[\mu}A_{\nu]} \ , \hskip 1 cm
\nabl_{[\mu}F_{\nu\rho]}= 0 .\eqno(2.4)$$
Although many other possibilities can in principle be enviseaged, for the
sake of simplicity the  following analysis will not go beyond allowance for
the only one that is important in applications to the kind of cosmic or
superfluid defects that are the subject of discussion in the present volume,
namely an antisymmetric Kalb-Ramond gauge field $B_{\mu\nu}=-B_{\nu\mu}$
whose exterior derivative is an automativally closed physical current
3-form,
$$N_{\mu\nu\rho}=3\nabl_{[\mu}B_{\nu\rho]} \ ,\hskip 1 cm
\nabl_{[\mu}N_{\nu\rho\sigma]}=0 .\eqno(2.5)$$
Just as a Maxwellian gauge transformation of the form $A_\mu\mapsto
A_\mu+\nabl_\mu\alpha$ for an arbitrary scalar $\alpha$ leaves the
electromagnetic field (2.4) invariant, so analogously a Kalb-Ramond gauge
transformation $B_{\mu\nu}\mapsto B_{\mu\nu}+2\nabl_{[\mu}\chi_{\nu]}$ for
an arbitrary covector $\chi_\mu$ leaves the corresponding current 3-form (5)
invariant. In applications to ordinary 4-dimensional spacetime, the current
3-form will just be the dual $N_{\mu\nu\rho}=\epsilon_{\mu\nu\rho\sigma}
N^\sigma$ of an ordinary current vector  $N^\mu$ satisfying a conservation
law of the usual type, $\nabl_\mu N^\mu=0$. Such a Kalb-Ramond
representation can be used to provide an elegant variational formulation for
ordinary perfect fluid theory\cite{[25]} and is particularly convenient for
setting up ``global" string models of vortices both in a simple cosmic
axion or Higgs field\cite{[26]}\cite{[27]}\cite{[28]} and in a 
superfluid\cite{[29]} such as liquid Helium-4.

In accordance with the preceeding considerations, the analysis that follows
will be based on the postulate that the action is covariantly and gauge
invariantly determined by specifying each scalar Lagrangian contribution
$\ud\ov\L$ as a function just of the background fields, $A_\mu$,
$B_{\mu\nu}$ and of course $g_{\mu\nu}$, and of any relevant internal fields
(which in the simplest non-trivial case -- exemplified by Witten\cite{[7]} 
type superconducting string models\cite{[30]}\cite{[31]} -- consist just of 
a phase scalar $\varphi$). In accordance with the restriction that the branes be
``pure" or ``fine" in the sense explained above, it is postulated that 
polarisation effects are excluded by ruling out couplings involving
gradients of the background fields. This means that the effect of making 
arbitrary infinitesimal ``Lagrangian" variations $d_{_{\rm L}}A_\mu$, 
$d_{_{\rm L}} B_{\mu\nu}$, $d_{_{\rm L}}g_{\mu\nu}$ of the background fields
will be to induce a corresponding variation $d\I$ of the action that simply
has the form
$$d\I=\sum_{\rm d}\int\!\! \ud d\!\ov{\cal S}\left\{
\ud\ov J{^\mu} d_{_{\rm L}} A_\mu +{_1\over ^2}\ud\ov W{^{\mu\nu}}
d_{_{\rm L}} B_{\mu\nu}+ {_1\over^2} \ud\ov T{^{\mu\nu}} d_{_{\rm L}}
 g_{\mu\nu} \right\} \ , \eqno(2.6)$$
provided either that that the relevant independent internal field components
are fixed or else that the internal dynamic equations of motion are
satisfied in accordance with the variational principle stipulating that
variations of the relevant independent field variables should make no
difference. For each d-brane of the complex, this partial differentiation
formula implicitly specifies the corresponding {\it electromagnetic surface
current density} vector $\ud\ov J{^{\mu}}$, the {\it surface vorticity flux} 
bivector $\ud \ov W{^{\mu\nu}}=-\ud\ov W{^{\nu\mu}}$, and the {\it surface stress 
momentum energy density} tensor $\ud\ov T{^{\mu\nu}}=\ud\ov T{^{\nu\mu}}$, 
which are formally expressible more explicitly as 
$$\ud\ov J{^{\mu}} = {\delta\!\ud \ov\L\over\delta A_\mu }
\ , \hskip 0.6 cm 
\ud\ov W{^{\mu\nu}} = 2{\delta\! \ud\ov\L\over\delta B_{\mu\nu}}
\ , \hskip 0.6 cm \ud\ov T{^{\mu\nu}}=2{\delta\!\ud \ov\L\over\delta 
g_{\mu\nu}}+\ud\ov\L\ud\eta^{\mu\nu}  
\ , \eqno(2.7)$$
where $\ud\eta^{\mu\nu}$ is the rank-d {\it fundamental tensor} of the
d-dimensional imbedding, as defined in the manner described in the
preceeding section.

The condition that the action be gauge invariant means that if one simply
sets $d_{_{\rm L}}A_\mu=\nabl_\mu \alpha$, $d_{_{\rm L}} B_{\mu\nu}
=2\nabl_{[\mu}\chi_{\nu]}$, $d_{_{\rm L}}g_{\mu\nu}=0$, for arbitrarily 
chosen $\alpha$ and $\chi_\mu$ then $d\I$ should simply vanish, i.e.
$$\sum_{\rm d}\int\!\! d\!\ud\ov{\cal S}\left\{
\ud\ov J{^\mu} \nabl_\mu\alpha+\ud\ov W{^{\mu\nu}}\nabl_\mu\chi_\nu
 \right\}=0 \ . \eqno(2.8)$$
In order for this to be able to hold for all possible fields $\alpha$ and
$\chi_\mu$ it is evident that the surface current $\ud\ov J{^\mu}$ and the
vorticity flux bivector $\ud\ov W{^{\mu\nu}}$ must (as one would anyway expect
from the consideration that they depend just on the relevant internal
d-surface fields) be purely d-surface tangential, i.e. their contractions
with the relevant rank (n--d) orthogonal projector $\ud\!\!\perp^\mu_{\
\nu}=g^\mu_{\ \nu}- \ud\eta^\mu_{\ \nu}$ must vanish:
$$\ud\!\!\perp^\mu_{\ \nu}\ud\ov J{^\nu}=0\ , \hskip 1 cm \ud\!\!
\perp^\mu_{\ \nu}\ud\ov W{^{\nu\rho}}=0\ .\eqno(2.9)$$
Hence, decomposing the full gradient operator $\nabl_\mu$ as the sum of
its tangentially projected part $\ud\ov\nabl_\mu=\ud\eta^\nu_{\ \mu}\nabl_\nu$
and of its orthogonally projected part $\ud\!\!\perp^\nu_{\ \mu}\nabl_\mu$, 
and noting that by (2.9) the latter will give no contribution, one sees 
that (2.8) will take the form
$$\sum_{\rm d}\int\!\!\ud d\!\ov{\cal S}\left\{ \ud\ov\nabl_\mu
\Big(\ud\ov J{^\mu} \alpha+\ud\ov W{^{\mu\nu}}\chi_\nu \Big)
-\alpha\ud\ov\nabl_\mu\ov J{^\mu} -\chi_\nu\ud\ov\nabl_\mu
\ud\ov W{^{\mu\nu}}\right\}=0  \ , \eqno(2.10)$$
in which first term of each integrand is a pure surface divergence.
Such a divergence can be dealt with using Green's theorem, according to which,
for any d-dimensional support surface $\ud\ov{\cal S}$ of a (d--1)-brane,
one has  the identity
$$\int\!\! \ud d\!\ov{\cal S}\ud\ov\nabl_\mu\ud\ov J{^\mu}=
\oint\! \udi d\!\ov{\cal S}\udp\lambda_\mu\ud\ov J{^\mu}\eqno(2.11)$$
where the integral on the right is taken over the boundary (d--1)-surface
of $\partial\ud\ov{\cal S}$ of $\ud\ov{\cal S}$, and $\udp\lambda_\mu$
is the (uniquely defined) unit tangent vector on the d-surface that is
directed normally outwards at its (d--1)-dimensional boundary. 
    Bearing in mind that a membrane support 
hypersurface can belong to the boundary of two distinct media,
and that for d$\leq$ n--3 a d-brane may belong to a common boundary
joining three or more distinct (d+1)-branes of the complex under 
consideration, one sees that (2.10) is equivalent to the condition
$$\sum_{\rm p}\int\!\! \up d\!\ov{\cal S}\,\Big\{ \alpha\Big(\!
\up\ov\nabl_\mu\up\ov J{^\mu}\! -\sum_{\rm d=p+1}\!\!\udp\lambda_\mu
\ud\ov J{^\mu}\Big)+ \chi_\nu\Big(\!\up\ov\nabl_\mu\up\ov W{^{\mu\nu}}\! 
-\sum_{\rm d=p+1}\!\!\udp\lambda_\mu\ud\ov W{^{\mu\nu}}\Big)\Big\}=0
 \ ,\eqno(2.12)$$
where, for a particular p-dimensionally supported (p--1)-brane,
the summation ``over d=p+1" is to be understood as consisting of a
contribution from each (p+1)-dimensionally supported p-brane attached to it,
where for each such p-brane, $\udp\lambda_\mu$ denotes the (uniquely defined)
unit tangent vector on its (p+1)-dimensional support surface that is
directed normally towards the p-dimensional support surface of the boundary
(p--1)-brane. The Maxwell gauge invariance requirement to the effect that
(2.12) should hold for arbitrary $\alpha$ can be seen to entail an
electromagnetic charge conservation law of the form
$$\up\ov\nabl_\mu\up\ov J{^\mu}=\sum_{\rm d=p+1}\!\!\udp\lambda_\mu
\ud\ov J{^\mu}\ . \eqno(2.13)$$
This can be seen from (2.11) to be be interpretable as meaning
that the total charge flowing out of particular (d--1)-brane from
its boundary is balanced by the total charge  flowing  into it from
any d-branes to which it may be attached. The analogous Kalb-Ramond
gauge invariance requirement that (2.12) should also hold for arbitrary
$\chi_\mu$ can be seen to entail a corresponding vorticity conservation
law of the form
$$\up\ov\nabl_\mu\up\ov W{^{\mu\nu}}=\sum_{\rm d=p+1}\!\!
\udp\lambda_\mu\ud\ov W{^{\mu\nu}}\ .\eqno(2.14)$$
A more sophisticated but less practical way of deriving the foregoing
conservation laws would be to work not from the expression (1) in terms  of
ordinary surface integrals but instead to use the superficially simpler
expression (2.2) in terms of distributions which leads to the replacement of
(2.13) by the ultimately equivalent (more formally obvious but less directly
meaningful) expression
$$\nabl_\mu\Big(\sum_{\rm d}\!\!\ud\hat J^{\mu}\Big)=0 \eqno(2.15)$$
involving the no longer regular but Dirac
distributional current $\ud\hat J^\mu$ that is given in terms of the
corresponding regular surface current $\ud\ov J^\mu$ by
$$ \ud \hat J^{\mu}=\Vert g \Vert^{-1/2}\int\!\!\ud d\!\ov{\cal S}\,
\ud \ov J{^{\mu}} \delta^{\rm n}[x-x\{\sigma\}]\ .\eqno(2.16) $$
Similarly one can if one wishes rewrite the vorticity flux conservation 
law (14) in the distributional form
$$\nabl_\mu\Big(\sum\!\! \ud \hat W^{\mu\nu}\Big)=0 \eqno(2.17)$$
where the distributional vorticity flux $\ud \hat W^{\mu\nu}$
is given in terms of the
corresponding regular surface flux $\ud\ov W^{\mu\nu}$ by
$$ \ud \hat W^{\mu\nu}=\Vert g \Vert^{-1/2}\int\!\!\ud d\!\ov{\cal S}\,
\ud \ov W{^{\mu\nu}} \delta^{\rm n}[x-x\{\sigma\}]\ .\eqno(2.18) $$
It is left as an entirely optional exercise for any  readers who may be
adept in distribution theory to show how the ordinary functional
relationships (2.13) and (2.14) can be recovered by by integrating out the 
Dirac distributions in (2.15) and(2.17).

The condition that the hypothetical variations introduced in (2.6) should be
``Lagrangian" simply means that they are to be understood to be measured
with respect to a reference system that is comoving with the various branes
under consideration, so that their localisation with respect to it remains
fixed. This condition is necessary for the variation to be meaningly
definable at all for a field whose support is confined to a particular brane
locus, but in the case of an unrestricted background field one can enviseage
the alternative possibility of an ``Eulerian" variation, meaning one defined
with respect to a reference system that is fixed in advance, independently
of the localisation of the brane complex, the standard example being that of
a Minkowski reference system in the case of a background that is flat. In
such a case the relation between the  more generally meaningfull Lagrangian
(comoving) variation, denoted by $d_{_{\rm L}}$, and the corresponding
Eulerian (fixed point) variation denoted by $d_{_{\rm E}}$ say will be given
by Lie differentiation with respect to the vector field $\xi^\mu$ say that
specifies the infinitesimal of the comoving reference system with respect to
the fixed background, i.e. one has
$$d_{_{\rm L}} -d_{_{\rm E}}=\vec{\ \xi\Libra}\eqno(2.19)$$
where the Lie differentiation operator $\vec{\ \xi\Libra}$ is given for the
background fields under consideration here by
$$\vec{\ \xi\Libra} A_\mu=\xi^\nu\nabl_\nu 
A_\mu+A_\nu\nabla_\mu\xi^\nu \ , $$ $$
\vec{\ \xi\Libra} B_{\mu\nu}=\xi^\nu\nabl_\nu B_{\mu\nu}
+2B_{\rho[\nu}\nabla_{\mu]}\xi^\rho \ , $$ $$
\vec{\ \xi\Libra} g_{\mu\nu}= 2\nabl_{(\mu}\xi_{\nu)}\ .\eqno(2.20)$$
 
This brings us to the main point of this section which is the derivation of
the dynamic equations governing the extrinsic motion of the branes of the
complex, which are obtained from the variational principle to the effect
that the action $\I$ is left invariant not only by infinitesimal variations
of the relevant independent intrinsic fields on the support suerfaces but
also by infinitesimal displacements of the support surfaces themselves.
Since the background fields $A_\mu$, $B_{\mu\nu}$, and $g_{\mu\nu}$ are to
be considered as fixed, the relevant Eulerian variations simply vanish, and
so the resulting Lagrangian variations will be directly identifiable with
the corresponding Lie derivatives -- as given by (2.20) -- with respect to the
generating vector field $\xi^\mu$ of the infinitesimal displacement under
consideration. The variational principle governing the equations of
extrinsic motion is thus obtained by setting to zero the result of
substituting these Lie derivatives in place of the corresponding Lagrangian
variations in the more general variation formula (2.6), which gives
$$ \sum_{\rm d}\int\!\! \ud d\!\ov{\cal S}\left\{
\ud\ov J{^\mu} \vec{\ \xi\Libra} A_\mu +{_1\over ^2}\ud\ov W{^{\mu\nu}}
\vec{\ \xi\Libra} B_{\mu\nu}+ {_1\over^2} \ud\ov T{^{\mu\nu}}\vec{\ \xi\Libra}
 g_{\mu\nu} \right\}=0 \ . \eqno(2.21)$$
The requirement that this should hold for any choice of $\xi^\mu$ evidently
implies that the tangentiality conditions (2.9) for the surface fluxes
 $\ud\ov J{^\mu}$ and $\ud\ov W{^{\mu\nu}}$ must be supplemented by an analogous
 d-surface tangentiality condition for the surface stress momentum energy 
tensor $\ud\ov T{^{\mu\nu}}$, which must satisfy
$$ \ud\!\!\perp^\mu_{\ \nu}\ud\ov T{^{\nu\rho}}=0\ .\eqno(2.22)$$
(as again one would  expect anyway from the consideration that it depends
just on the relevant internal d-surface fields).
This allows  (2.20) to be written out in the form
 $$ \sum_{\rm d}\int\!\!\ud d\!\ov{\cal S}\Big\{
\xi^\rho\Big(F_{\rho\mu}\ud\ov J{^\mu}\!+{_1\over^2}N_{\rho\mu\nu}\ud\ov W
^{\mu\nu}\! -\ud\ov\nabl_\mu\ud\ov T{^\mu_\rho}-A_\rho\ud\ov\nabl_\mu
\ud\ov J{^\mu}\! -B_{\rho\nu}\ud\ov\nabl_\mu\ud\ov W{^{\mu\nu}}\Big) $$
$$ \hskip 2 cm + \ud\ov\nabl_\mu\Big(\xi^\rho(
A_\rho \ud\ov J{^\mu}\! + B_{\rho\nu} \ud\ov W{^{\mu\nu}}\!+\ud\ov 
T{^\mu}_\rho)\Big) \Big\}=0 \ , \eqno(2.23)$$
in which the final contribution is a pure surface divergence that can be
dealt with using Green's theorem as before. Using the results (2.13) and 
(2.14) of the analysis of the consequences of gauge invariance and 
proceeding as in their derivation above, one sees that the condition for 
(2.23) to hold for an arbitrary field $\xi^\mu$ is that, on each (p--1)-brane
of the complex, the dynamical equations
$$\up\ov\nabl_\mu\up\ov T{^\mu}_\rho= \up f_\rho \ ,\hskip 1 cm
\up f_\rho=\up\ov f_\rho+\up\oc f_\rho\eqno(2.24)$$
should be satisfied, where the contact force density $\up\oc f_\rho$ 
exerted on the p-surface by other members of the brane complex is 
expressible as
$$\up\oc f_\rho= \sum_{\rm d=p+1}\!\!
\udp\lambda_\mu\ud\ov T{^\mu}_\rho  \ ,\eqno(2.25)$$
while the surface force density $\up\ov f_\rho$
exerted by the background fields is given by
$$\up\ov f_\rho=F_{\rho\mu}\up\ov J{^\mu}+{_1\over^2}
N_{\rho\mu\nu}\up\ov W{^{\mu\nu}}\ .\eqno(2.26)$$
 As before the summation ``over d=p+1" in (2.25) is to be understood as 
consisting of a contribution from each of the p-branes attached to 
the (p--1)- brane under consideration, where for each such attached p-brane,
$\udp\lambda_\mu$ denotes the (uniquely defined) unit tangent vector on its
(p+1)-dimensional support surface that is directed normally towards the
p-dimensional support surface of the boundary (p--1)-brane. The first of
the background force contibutions in (2.26) is of course the Lorentz type
force density resulting from the effect of the electromagnetic field on
 the surface current, while the other contribution in (2.26) is a Joukowsky 
type force density (of the kind responsible for the lift on an aerofoil) 
resulting from the Magnus effect, which acts in the case of a ``global" 
string\cite{[26]}\cite{[27]} though not in the case of a string of the 
``local" type for which the relevant vorticity flux $\up\ov W{^{\mu\nu}}$
will be zero. As with the conservation laws (2.13) and (2.14), so also the
explicit force density balance law expressed by (2.24)  can alternatively be
expressed in terms of the corresponding Dirac distributional stress momentum
energy and background force density tensors, $\ud\hat T^{\mu\nu}$ and
$\ud\hat f_\mu$, which are given for each (d--1)-brane in terms of the
corresponding regular surface stress momentum energy and background force
density tensors $\ud\ov T{^{\mu\nu}}$ and $\ud\ov f_\mu$ by
$$ \ud \hat T^{\mu\nu}=\Vert g \Vert^{-1/2}\int\!\!\ud d\!\ov{\cal S}\,
\ud \ov T{^{\mu\nu}} \delta^{\rm n}[x-x\{\sigma\}]\ \eqno(2.27) $$
and
$$ \ud \hat f_\mu=\Vert g \Vert^{-1/2}\int\!\!\ud d\!\ov{\cal S}\,
\ud \ov f_\mu \delta^{\rm n}[x-x\{\sigma\}]\ . \eqno(2.28) $$
The equivalent -- more formally obvious but less explicitly meaningful --
distributional versional version of the force balance law (2.24) takes 
the form
$$\nabl_\mu\Big(\sum_{\rm d}\!\! \ud \hat T^{\mu\nu}\Big)
=\hat f_\mu \eqno(2.29)$$
where the total Dirac distributional force density is given in terms of the 
electromagnetic current distributions (2.16) and the vorticity flux 
distributions (2.18) by
$$\hat f_\mu=F_{\rho\mu}\sum_{\rm d}\!\!\ud\hat J^\mu+{_1\over^2}
N_{\rho\mu\nu}\sum_{\rm d}\!\!\ud\hat W^{\mu\nu}\ .\eqno(2.30)$$
It is again left as an optional exercise for readers who are adept in the
use of Dirac distributions to show that the system (2.24), (2.25), (2.26)
 is obtainable from
(2.29) and (2.30) by substituting (2.16), (2.18), (2.27), (2.28).

As an immediate corollary of (2.24), it is to be noted that for any 
vector field $k^\mu$ that generates a continuous symmetry of the background
spacetime metric, i.e. for any solution of the Killing equations
$$\nabl_{(\mu}k_{\nu)}=0\ ,\eqno(2.31)$$
one can construct a corresponding surface momentum or energy density current
$$\up\ov{P}{^\mu}=\up\ov T{^{\mu\nu}}k_\nu\ ,\eqno(2.32)$$
that will satisfy
$$\up\ov\nabl_\mu\up\ov{P}{^\mu}=
 \sum_{\rm d=p+1}\!\!
\udp\lambda_\mu\ud\ov{P}{^\mu}+\up\ov f_\mu k^\mu  \ .\eqno(2.33)$$
In typical applications for which the n-dimensional
background spacetime can be taken to be flat there will be n independent
translation Killing vectors which alone (without recourse to the further
n(n--1)/2 rotation and boost Killing vectors of the Lorentz algebra)
 will provide a set of relations of the form (2.33) that together
provide the same information as that in the full force balance
equation (2.24) or (2.29).

Rather than the distributional version (2.29), it is the explicit version 
(2.24) of the force balance law that is directly useful for calculating 
the dynamic evolution of the brane support surfaces. Since the relation
(2.29) involves n independent components whereas the support surface involved
is only p-dimensional, there is a certain redundancy, which results from
the fact that if the virtual displacement field $\xi^\mu$ is 
tangential to the surface in question it cannot affect the action. Thus
if $\up\!\!\perp^{\!\mu}_{\,\nu}\xi^\nu=0$, the condition (2.21) will be 
satisfied as a mere identity -- provided of course that the field equations
governing the internal fields of the system are satisfied. It follows that
the non-redundent information governing the extrinsic motion of the
p-dimensional support surface will be given just by the orthogonally
projected part of (2.24). Integrating by parts, using the fact that, 
by (1.6) and (1.15), the surface gradient of the rank-(n--p) orthogonal 
projector $\up\!\!\perp^{\!\mu}_{\,\nu}$ will be given in terms of the second 
fundamental tensor $\up K_{\mu\nu}^{\ \ \,\rho}$ of the p-surface by
$$\up\ov\nabl_\mu\up\!\!\perp^{\!\nu}_{\,\rho}=-\up K_{\mu\nu}^{\ \ \,\rho}
-\up K_{\mu\ \nu}^{\ \rho} \ ,\eqno(2.34)$$
it can be seen that the extrinsic equations of motion obtained as the
orthogonally projected part of (2.24) will finally be expressible by
$$\up\ov T{^{\mu\nu}}\up K_{\mu\nu}^{\ \ \,\rho}=\up\!\!\perp^{\!\rho}_{\,\mu}
\up f^\mu\ .\eqno(2.35)$$

\def\spose#1{\hbox to 0pt{#1\hss}} 
\def\Libra{\spose {--} {\cal L}}
\def\Diam{\spose {\raise 0.3pt\hbox{+}} {\diamondsuit}  }

\def\eqdef{\fff\ \vbox{\hbox{$_{_{\rm def}}$} \hbox{$=$} }\ \thf }
\def\ov{\overline}
\def\oov{\offinterlineskip\overline}
\def\oc{\check}
\def\ot{\widetilde}
\def\dL{d_{_{\rm L}}}

\def\uth{{^{\,_{(3)}}}\!}  \def\utw{{^{\,_{(2)}}}\!}
\def\ud{{^{\,_{(\rm d)}}}\!}  \def\udi{{^{\,_{(\rm d-1)}}}\!}
\def\up{{^{\,_{(\rm p)}}}\!}  \def\udp{{^{\,_{(\rm d)}}}\!}

\def\og{\eta}

\def\nabl{\nabla\!} \def\onab{\ov\nabl}

\def\vv{_{\ \vert}}     \def\osb{\offinterlineskip\hbox}
\def\af{\fff\vbox{\osb{$_{_{\vv}}$} \hbox{$\ov {\ f} $} }\thf }
\def\ag{\perp\!}

\def\P{{\cal P}} \def\R{{\cal R}} \def\W{{\cal W}} \def\E{{\cal E}} 
\def\I{{\cal I}} \def\L{{\cal L}}
\def\A{{_A}} \def\B{{_B}} \def\X{{_X}} \def\Y{{_Y}}

\def\lag{{\cal L}}
\def\aag{{m^{\rm d}}} \def\bag{{\rm b}}   \def\cag{{\rm c}}

\def\pag{{\cal P}} \def\kag{{\mit\Sigma}}
\def\j{J}    \def\w{W} \def\c{\chi}

It is to be emphasised that the formal validity of the formula that has just 
been derived is not confined to the variational models on which the above 
derivation is based, but also extends to dissipative models (involving 
effects such as external drag by the background 
medium\cite{[12]}\cite{[13]}\cite{[32]} or
mutual resistence between independent internal currents). The condition
that even a non-conservative macroscopic model should be compatible with
an underlying microscopic model of conservative type  requires
the existence (representing to averages of corresponding microscopic 
quantities) of appropriate stress momentum energy density and force
density fields satisfying (2.35). This ubiquitously applicable 
formula is just the natural higher generalisation 
of ``Newton's law" (equating the product of mass with acceleration to the 
applied force) in the case of a particle. The surface stress momentum 
energy tensor, $\up\ov T{^{\mu\nu}}$, generalises the mass,
and the second fundamental tensor, $\up K_{\mu\nu}^{\ \ \,\rho}$, 
generalises the acceleration. This can be seen from the fact that 
in the 1-dimensional case of a ``pure" point particle (i.e. a monoplole) 
of mass $m$, the Lagrangian is given simply by $^{_{(1)}}\!\ov\L=-m$,
 so the 1-dimensional energy tensor is obtained in terms
of the unit tangent vector $u^\mu$ ($u^\mu u_\mu=-1$) as $^{_{(1)}}\!
\ov T{^{\mu\nu}}=mu^\mu u^\nu$. In this zero-brane case, the first fundamental 
tensor is simply given by $^{_{(1)}}\!\eta^{\mu\nu}=-u^\mu u^\nu$, and the
second fundamental tensor is given in terms of the acceleration $\dot u^\mu
=u^\nu\nabl_\nu u^\mu$ as $^{_{(1)}}\!K_{\mu\nu}^{\ \ \,\rho}=u_\mu u_\nu
\dot u^\rho$. Thus (2.35) can be seen to reduce in the case of a particle
simply to the usual familiar form
$m\dot u^\rho=^{_{(1)}}\!\perp^{\!\rho}_{\,\mu}{^{_{(1)}}}\!f^\mu$.

\section{Perturbations and Curvature Effects beyond the Pure Brane Limit}

Two of the most useful formulae for the analysis of small perturbations
of a string or higher brane worldsheet are the expressions for the
infinitesimal Lagrangian (comoving) variation of the first and
second fundamental tensors in terms of the corresponding
 comovings variation $\dL g_{\mu\nu}$ of the metric (with respect
to the comoving reference system). For the first fundamental
tensor one easily obtains 
$$\dL\og^{\mu\nu}= -\og^{\mu\rho}\og^{\mu\sigma}\dL g_{\rho\sigma}\ ,
\hskip 1 cm\dL \og^\mu_{\ \nu}=\og^{\mu\rho}\!\perp^{\!\sigma}_{\,\nu}
\dL g_{\rho\sigma} 
\eqno(3.1)$$
and, by substituting this in the defining relation (1.10),  
the corresponding Lagrangian variation of the second 
fundamental tensor is obtained\cite{[33]} as 
$$\dL K_{\mu\nu}^{\ \ \,\rho}=\perp^{\!\rho}_{\,\lambda}\og^\sigma_{\,\mu}
\og^\tau_{\,\nu}\,\dL\Gamma_{\sigma\ \tau}^{\ \lambda} +\big(2\perp^{\!\sigma}
_{\,(\mu} K_{\nu)}^{\ \ \tau\rho}-K_{\mu\nu}^{\ \ \,\sigma}\og^{\tau\rho}
\big)\dL g_{\sigma\tau} \ ,\eqno(3.2)$$
where the Lagrangian variation of the connection (1.18) is given 
by the well known formula
$$\dL\Gamma_{\sigma\ \tau}^{\ \lambda} =g^{\lambda\rho}\big(
\nabl_{(\sigma\,}\dL g_{\tau)\rho}-{_1\over^2}\nabl_{\rho\,}\dL g_{\sigma\tau}
\big)\ .\eqno(3.3)$$
Since we are concerned here only with cases for which the background
is fixed in advance so that the Eulerian variation
$d_{\rm_E}$ will vanish in (2.19), the Lagrangian variation of the metric
will be given just by its Lie derivative with respect to the infinitesimal
displacement vector field $\xi^\mu$ that generates the displacement of the
worldsheet under consideration, i.e. we shall simply have
$$\dL g_{\sigma\tau}=2\nabl_{(\sigma}\xi_{\tau)} \ .\eqno(3.4)$$
It then follows from (3.3) that the Lagrangian variation of the connection
will be given by
$$\dL\Gamma_{\sigma\ \tau}^{\ \lambda}=\nabl_{(\sigma}\nabl_{\tau)}
\xi^\lambda-\R^\lambda_{\ (\sigma\tau)\rho}\xi^\rho\ ,\eqno(3.5)$$
where $\R^\lambda_{\ \sigma\tau\rho}$ is the background Riemann curvature 
(which will be negligible in typical applications for which the lengthscales
characterising the geometric features of interest will be small compared
with those characterising any background spacetime curvature).
The Lagrangian variation of the first fundamental tensor is thus
finally obtained in the form
$$\dL\og^{\mu\nu}=-2\og_\sigma^{\,(\mu}\ov\nabla{^{\nu)}}\xi^\sigma\ ,
\eqno(3.6)$$
while that of the second fundamental tensor is found to be given by
$$\dL K_{\mu\nu}^{\ \ \,\rho}=\perp^{\!\rho}_{\,\lambda}\big(
\ov\nabl_{(\mu}\ov\nabl_{\nu)}\xi^\lambda-\og^\sigma_{\,(\mu}
\og^\tau_{\ \nu)}\R^\lambda_{\ \sigma\tau\rho}\xi^\rho-K^\sigma_{\ (\mu\nu)}
\ov\nabl_\sigma\xi^\lambda\big)+$$ $$\hskip 2cm
\big(2\perp^{\!\sigma}_{\,(\mu}K_{\nu)\tau}^{\ \ \ \rho}-g^\rho_{\,\tau}
K_{\mu\nu}^{\ \ \,\sigma}\big)\big(\nabl_\sigma\xi^\tau +
\ov\nabla{^\tau}\xi_\sigma\big)\  .\eqno(3.7)$$

It is instructive to apply the forgoing formulae to the case of a
{\it free} pure brane worldsheet, meaning one for which there is no external
force contribution so that the equation of extinsic motion reduces to the
form
$$\ov T{^{\mu\nu}}K_{\mu\nu}^{\ \ \,\rho}=0\ .\eqno(3.8)$$
On varying the relation (3.8) using (3.7) in conjunction
with the orthogonality property (2.22) and the unperturbed equation (3.8)
itself, the equation governing the propagation of the infinitesimal
displacement vector is obtained in the form
$$\perp^{\!\rho}_{\, \lambda}\ov T{^{\mu\nu}}\big(\ov\nabl_\mu\ov\nabl_\nu
\xi^\lambda-\R^\lambda_{\ \mu\nu\sigma}\xi^\sigma\big)
=-K_{\mu\nu}^{\ \ \,\rho\,}\dL \ov T{^{\mu\nu}}\ .\eqno(3.9)$$

The extrinsic perturbation equation (3.9) is by itself only part of the
complete system of perturbation equations governing the evolution of the
brane, the remaining equations of the system being those governing the
evolution of whatever surface current\cite{[36]}and other relevant internal
fields on the supporting worldsheet may be relevant. The perturbations
of such fields are involved in the source term on
the right of (3.9), whose explicit evaluation depends on the specific form
of the relevant currents or other internal fields. However it is not
necessary to know the specific form of such internal fields for the purpose
just of deriving the characteristic velocities of propagation of the
extrinsic propagations represented by the displacement vector $\xi^\mu$, so
long as they contribute to the source term on the right of the linearised
perturbation equation (3.9) only at first differential order, so that the
characteristic velocities will be completely determined by the first term on
the left of (3.9) which will be the only second differential order
contribution. It is apparent from (3.9) that under these conditions the
equation for the characteristic tangent covector $\chi_\mu$ say will be
given independently of any details of the surface currents or other internal
fields simply\cite{[4]} by
$$\ov T{^{\mu\nu}}\chi_\mu\chi_\nu=0\ .\eqno(3.10)$$
(It can be seen that the unperturbed surface stress momentum energy density
tensor $\ov T{^{\mu\nu}}$ plays the same role here as that of the unperturbed
metric tensor $g^{\mu\nu}$ in the analogous characteristic equation for the
familiar case of a massless background spacetime field, as exemplified by
electromagnetic or gravitational radiation.)

The linerised perturbation equation (3.9) will of course completely 
determine the evolution of the displacements by itself if there are no 
internal fields, i.e. in the case of the Dirac Gotu Nambu model determined 
by a surface Lagrangian of the trivial constant form
$$\ud\ov \lag=-m^{\rm d} \eqno(3.11)$$
where $m$ is a constant having the dimension of mass (which would be of the
order of magnitude of the relevant Higgs mass scale in the case of a vacuum
defect arising from the spontaneous symmetry breaking mechanism of the kind
most commonly considered\cite{[5]}) and d is the dimension of the worldsheet
(i.e. d=1 for a simple point particle, d=2 for a string, and so on). In
this rather degenerate special case the surface stress momentum energy 
density tensor determined according to the prescription (2.7) will be
given simply by
$$\ud\ov T{^{\mu\nu}}=-m^{\rm d}\ud\og{^{\mu\nu}}\ .\eqno(3.12)$$
The unperturbed Dirac Gotu Nambu equation of motion is thus obtained in a 
form that is given -- independently of the dimension d indicated by the 
prefix $^{(\rm d)}$ which may therefore be dropped -- by the well known
harmonicity condition that is expressible as the vanishing of the curvature
vector,
$$K^\mu=0\ .\eqno(3.13)$$
The corresponding perturbation equation is obtained from
(3.9) in the form
$$\perp^{\!\rho}_{\, \lambda}\big(\ov\nabla^\mu\ov\nabl_\mu
\xi^\lambda-\og^{\mu\nu}\R^\lambda_{\ \mu\nu\sigma}\xi^\sigma\big)
=2K_{\mu\nu}^{\ \ \,\rho\,}\ov\nabla^\mu\xi^\nu\ .\eqno(3.14)$$
As well as the compact tensorial version\cite{[33]} given here, the 
litterature includes other equivalent but formally more complicated 
expressions\cite{[59]}\cite{[34]} (involving reference to internal coordinates or 
surface adapted frames of the kind discussed at the beginning of Section 1)
that generalise earlier work restricted to the hypersurface supported
(``wall" or membrane) special 
case\cite{[60]}\cite{[61]}\cite{[62]}\cite{[63]}.

 For most practical physical purposes the most useful generalisations of the
Dirac-Goto-Nambu models governed by (3.12) and  (3.13) are those of the very
general category governed by (3.8) and (3.9) which allow for internal fields
such as the currents that can be used, not only to represent the Witten type
superconductivity effect in cosmic strings, but also also to represent the
effect of ordinary elasticity in terrestrial applications, such as the
strings of musical instruments which were already a subject of scientific
investigation, albeit at an empirical rather than theoretical level, in the
time of Pythagoras. However before proceeding to the discussion of such
internal field effects in the following section, it is of interest to
consider how the simple Dirac-Goto-Nambu model can be generalised in a way
that goes beyond the ``pure" brane description characterised in the free
case by (3.8) and in the presence of an external force by (2.35). The
distinguishing property of a ``pure" brane model of the kind considered in
the previous section is the condition that the action depends only on the
undifferentiated background fields $g_{\mu\nu}$, $A_\mu$, $B_{\mu\nu}$, but
not on their gradients. One of the most familiar kinds of example in which
this condition fails to hold is that of an electromagnetically polarised
medium, whose action\cite{[35]} depends not just on the gauge field $A_\mu$ but
also directly on the associated field $F_{\mu\nu}$ itself. However for lack
of time and space we shall not consider such electromagnetic effects in the
present section but will consider only the simplest category of
``geodynamic" brane models, meaning those in which, as in the ``pure" brane
models of  the Dirac-Goto-Nambu category, the action depends only on the
imbedding geometry of the worldsheet and not on any other external internal
fields. The simplest such extension of the ``pure" Dirac-Goto-Nambu model
(whose action is proportional just to the surface measure which depends only
on $g_{\mu\nu}$ but not its derivatives) is based on a Lagrangian consisting
not only of a constant term but also of terms proportional to the two
independent scalars that can be constructed as quadratic functions of the
first derivatives of the metric, namely $K_\mu K^\mu$ and
$K_{\mu\nu\rho}K^{\mu\nu\rho}$. The inclusion of such ``stiffness" terms has
been suggested by Polyakov and others\cite{[37]}, one of the main reasons
being allowance for the deviations from the ``pure" Dirac-Goto-Nambu 
description of cosmological string\cite{[38]}\cite{[39]}\cite{[40]} or domain 
wall\cite{[41]}\cite{[42]}\cite{[43]}\cite{[44]}\cite{[45]} defects that one
would expect to arise if the curvature becomes too strong. The kind of 
Lagrangian constructed in this way, namely
$$\ud\ov\lag=-\aag +\bag\,\ud K_\rho\ud K^\rho-\cag\,\ud K_{\mu\nu\rho}
\ud K^{\mu\nu\rho}\ ,\eqno(3.15)$$
where $m$, $\bag$, and $\cag$ are constants, has recently been the subject of 
several investigations\cite{[46]}\cite{[47]}\cite{[120]}\cite{[48]}\cite{[49]}.
It is convenient to use the abbreviation
$$\ud\kag^{\mu\nu\rho}=\bag\,\ud\og^{\mu\nu}\ud K^\rho-\cag\,
\ud K^{\mu\nu\rho}\,\eqno(3.16)$$
which enables the Lagrangian (3.15) to be expressed in condensed form as
$$\ud\ov\lag=-\aag+\ud\kag_{\mu\nu\rho}\ud K^{\mu\nu\rho} \ .\eqno(3.17)$$
The variation needed for evaluating the change in such an ``impure"
brane action will thereby be obtainable from the formulae above in the 
corresponding form
$$\Vert h \Vert^{-1/2}\dL\big(\Vert h \Vert^{1/2}\ov\lag\big)
=\dL\ov\lag+{1\over 2}\ov\lag\,\og^{\mu\nu}\dL g_{\mu\nu}\ \eqno(3.18) $$  
(again dropping the explicit reference to the brane dimension d) with
$$\dL\ov\lag=\big(\kag^{\lambda\rho\mu} K_{\lambda\rho}^{\ \ \nu}
-2\,\kag^{\lambda\mu}_{\ \ \rho} K_\lambda^{\ \nu\rho}\big)\dL g_{\mu\nu}
+\big(2\,\kag^{\mu\lambda\nu}-\kag^{\mu\nu\lambda}\big)
\nabl_\lambda\dL g_{\mu\nu} \ .\eqno(3.19)$$
Although it is still possible to construct a formally symmetric stress
momentum energy density tensor of the distributional type, the presence of
the gradient term on the right of (3.19) will make it rather pathological,
with not just a contribution proportional to a Dirac distribution as in the
``pure" brane case described by (2.27) but also with a contribution
proportional to the even more highly singular gradient of a Dirac
distribution\cite{[48]}. In order to be able to continue working in terms of
strictly regular surface supported field, it is necessary\cite{[49]} to deal
with the gradient dependence of the action in such an ``impure" brane model 
by having recourse to the use of a total stress momentum energy tensor ${\cal
T}^\mu{_{\nu}}$ of the no longer no longer symmetric canonical type that
 can be read out from the variation formula
$$\Vert h \Vert^{-1/2}\dL\big(\Vert h \Vert^{1/2}\ov\lag\big)
={\cal T}^\mu{_{\nu}}\ov\nabl_\mu\xi^\nu+2\kag_\mu^{\ 
\rho\sigma}\,{\cal R}_{\rho\sigma\ \nu}^{\ \ \,\mu}\xi^\nu 
-\,\ov\nabl_\mu\big(2\,\kag_\nu^{\ 
\mu\rho}\nabl_\rho\xi^\nu\big)\eqno(3.20) $$
that is obtained after substitution of (3.4) in (3.18) and (3.19). The
regular but non-symmetric {\it canonical surface stress momentum energy
density} tensor ${\cal T}^\mu{_{\nu}}$ obtained in this way\cite{[49]}
is given by 
$${\cal T}^\mu{_{\nu}}=\ot T^\mu_{\ \,\nu}+t^\mu_{\ \nu}\ , \eqno(3.21)$$
where $\ot T^{\mu\nu}$ is symmetric and purely tangential to the worldsheet
(as a regular ``geometric" stress momentum energy density tensor would be)
with a merely algebraic dependence on the second fundamental tensor, being
given by
$$\ot T^\mu{_{\nu}}=\ot T_\nu^{\ \mu}={\cal T}^\mu{_{\!\lambda}}
\og^\lambda_{\ \nu}= \ov\lag\,\og^\mu_{\ \nu}-2\,\kag^\lambda_{\ \nu\rho}\,
K_\lambda^{\ \mu\rho}\ ,\eqno(3.22)$$
while the remainder, which is of higher differential order, is expressible in
 terms of the third
fundamental tensor (1.22) as
$$t^\mu_{\ \nu}={\cal T}^\mu{_{\!\lambda}}\!\perp^{\!\lambda}_{\,\nu}=
2\,\cag\,\Xi^{\lambda\ \mu}_{\ \,\lambda\ \nu}
- 2\,{\bag}\,\Xi^{\mu\ \lambda}_{\ \,\lambda\ \nu} \ , \eqno(3.23)$$
which can conveniently be rewritten with the higher derivative
contributions regrouped in the form
$$t^\mu_{\ \nu}=2(\cag-\bag)\Xi^\mu_{\ \nu}+\cag\,\og^\mu_{\ \rho}
\og^{\sigma\tau}\R^\rho{_{\sigma\tau\lambda}}\ag^{\!\lambda}_{\,\nu}
\ ,\eqno(3.24)$$
where
$$\Xi^\mu_{\ \nu}=\Xi^{\mu\ \lambda}_{\ \lambda\ \nu}=\ag^{\!\lambda}_{\,\nu}
\ov\nabla^\mu K_\lambda \ .\eqno(3.25)$$
It is to be noticed that the total canonical surface stress momentum energy 
density tensor obtained in this way is is still automatically 
tangential to the worldshheet on its first (though no longer
on its second) index, i.e.
$$\perp^{\!\lambda}_{\,\mu}\!{\cal T}^\mu{_{\nu}}=0\ ,\eqno(3.26)$$
and that the higher derivative contribution proportional to the trace
$\Xi^\mu_{\ \nu}$ of the third fundamental tensor will drop out if the
coefficients have the same value, $\cag=\bag$.

The application of the variation principle to the effect that the surface
integral of the variation (3.20) should vanish for any displacement
$\xi^\mu$ within a bounded neighbourhood can be seen  to lead
(via an application of Green's theorem as in the preceeding section)
to dynamical equations\cite{[49]} of the form 
$$\ov\nabl_\mu{\cal T}^\mu{_{\!\nu}}=2\,\kag_\mu^{\ \rho\sigma}\,
{\cal R}_{\rho\sigma\ \nu}^{\ \ \,\mu} \ .\eqno(3.27)$$
As in the pure Dirac-Gotu-Nambu case discussed above, the foregoing system 
of equations is partially redundant: although it involves n distinct 
spacetime vectorial equations, only n--d of them are dynamically 
independent, namely those projected orthogonally to the d-dimensional 
worldsheet. The others are merely Noether identities which follow 
independently of the variation principle from the fact that a 
displacement $\xi^\nu$ that is purely tangential to the worldsheet merely 
maps it onto itself and thus cannot affect the action, as can be verified 
directly using the generalised Codazzi identity (1.23).  

The lack of symmetry of ${\cal T}^\mu{_{\!\nu}}$ means that the construction
of the corresponding momentum current vector, ${\cal P}^\mu$ say, associated
with a generic background spacetime Killing vector field $k^\mu$ will
not be quite as simple as in the ``pure" brane case for which an expression
of the simple form (2.32) suffices. However using the well known
fact that the Killing equation (2.31) entails the integrability condition
$$\nabl_\mu\nabl_\nu k^\rho=\R^\rho_{\ \nu\rho\lambda}k^\lambda\ ,
\eqno(3.28)$$
together with the observation that the antisymmetric part of the canonical 
stress momentum energy density tensor (3.22) is given according to (3.23) 
just by
$${\cal T}^{[\mu\nu]}=-2\ov\nabl_\lambda\Sigma^{\lambda[\mu\nu]}
\ ,\eqno(3.29)$$
it can be seen that, for any solution of (2.31), the ansatz\cite{[49]}
$${\cal P}^\mu={\cal T}^\mu{_{\!\nu}}k^\nu+ 2\Sigma^{\mu\nu}{_\rho}
\nabl_\nu k^\rho \eqno(3.30)$$
provides a surface current, ${\cal P}^\mu$, which satisfies the 
tangentiality condition
$$\perp^{\!\lambda}_{\,\mu}{\cal P}^\mu=0\ ,\eqno(3.31)$$
and for which the strict surface conservation law,
$$\ov\nabl_\mu{\cal P}^\mu=0\ , \eqno(3.30)$$  
will hold whenever the equation of motion (3.25) is satisfied.
It is to be remarked however that for a Killing vector of the irrotational
kind for which $\nabl_\mu k_\mu$ vanishes altogether the second term
(interpretable as a surface spin density contribution) in (3.30) will not
contribute, i.e. an expression of the simpler form (2.32) will suffice.
This applies in particular to the case of an ordinary translation 
generator in flat space, for which the corresponding conserved surface current
will represent ordinary energy or linear momentum, whereas in the case of
angular momentum the extra (spin density) term in (2.32) is indispensible.

The preceeding formulae all include allowance for arbitrary background 
 curvature, but, to obtain the analogue of the non redundant
version (3.8) of the equations of motion in a reasonably simple form,
the restriction that the background spacetime be flat, i.e.
$\R^\mu{_{\nu\rho\sigma}}=0$, will now be imposed. This enables
the required system of dynamical equations to be expressed\cite{[49]} in 
the form
$$\ot T{^{\mu\nu}}K_{\mu\nu\rho}=2(\bag-\cag) 
\ag^{\!\nu}_{\,\rho}\ov\nabl_\mu\big(\ag^{\!\sigma}_{\,\nu}\ov\nabla^\mu
K_\sigma\big)\ ,\eqno(3.31)$$
with the higher derivative terms grouped on the right hand side, which
vanishes if $\bag=\cag$.

In the particular case\cite{[45]} of a membrane, meaning a brane supported 
by a hypersurface of dimension d=n--1, the second fundamental tensor and its
trace will be given in terms of the unit normal $\lambda_\rho$ (which
in this case will be unique up to a choice of sign) by
$K_{\mu\nu\rho}=K_{\mu\nu}\lambda_\rho$ and $K_\rho= K\lambda_\rho$
with $K=K^\mu_{\ \mu}$ where $K_{\mu\nu}$ is the second fundamental
form (whose sign depends on that orientation chosen for the normal).
In an ordinary 4-dimensional spacetime background, this membrane case
corresponds to d=3, for which the symmetric tangential
part of the surface stress momentum energy density tensor will be
therefore be obtainable from (3.22) in the form
$$\uth\ot T{^{\mu\nu}}=-\big(m^3-\bag\uth K^2+\cag\uth K^\rho_{\ \sigma}
\uth K^\sigma_{\ \rho}\big)\uth\og^{\mu\nu}-2\bag\uth K\uth K^{\mu\nu}
+2\cag \uth K^{\lambda\mu}\uth K_\lambda^{\ \nu}\ .$$
In the case of a string with d=2 one can use the fact that the
trace free conformation tensor (1.20) will satisfy
$2\utw C^\lambda_{\  \,\mu\rho}\utw C_{\lambda\nu}^{\ \ \,\rho}
=$ $\utw C_{\kappa\lambda\rho}\utw C^{\kappa\lambda\rho}\utw\og_{\mu\nu}$
to obtain a corresponding formula (which holds regardless of the background
spacetime dimension n) given\cite{[49]} by
$$\utw\ot T{^{\mu\nu}}=-m^2\utw\og^{\mu\nu}+2(\cag-\bag)\utw C^{\mu\nu\rho}
\utw K_\rho \ .$$

\section{Conservative String Models as examples of perfect Branes}

It is reasonable to postulate that a ``weak" energy condition of the kind
formulated and justified by Hawking and Ellis\cite{[50]} should hold for any
pure p-brane model as a  condition for physical realism as a macroscopic
description of a (p+1)-surface supported physical system at a classical
level, meaning that the model's surface stress momentum energy
density tensor (as introduced in section 2) should be such that 
the contraction $\ov T{^{\mu\nu}}\beta_\mu\beta_\nu$ is non negative for
any vector $\beta^\mu$ that is timelike. Furthermore the causality condition
to the effect that there should be no timelike characteristic covector 
(i.e. no superluminal propagation) can be seen from (3.10) to entail the 
further requirement (going marginally beyond the ``weak" condition of 
Hawking and Ellis) that $\ov T{^{\mu\nu}}\beta_\mu\beta_\nu$ should
be strictly positive if $\beta^\mu$ is timelike. This leads to the 
formulation of what may be called the ``minimal" energy condition 
for a pure p-brane which is expressible  as
$$\beta^\mu\beta_\mu<0\hskip 0.6 cm \Rightarrow\hskip 0.6 cm
 \ov T{^{\mu\nu}}\beta_\mu\beta_\nu>0 \ .\eqno(4.1)$$

A (pure) p-brane model will consequently be characterised by a well defined  
{\it surface energy density}, $U$ say,
 that is specifiable by an eigenvalue equation of the form 
$$\ov T{^\mu}{_\nu}\beta^\nu=-U \beta^\mu \eqno (4.2) $$ 
where the corresponding eigenvector $\beta^\mu$ is distinguished by the
requirement that it be tangential and non-spacelike:
$$\ag^{\!\mu}_{\,\nu}\beta^\nu=0\ , \hskip 1 cm \beta^\mu\beta_\mu
\leq 0 \ .\eqno(4.3)$$
It is apparent that the ``minimal''energy condition (4.1) requires
that the eigenvalue $U$ should be strictly positive unless $\beta^\mu$
is null in which case it may vanish:
$$U\geq 0\ ,\hskip 1 cm U=0\hskip 0.6 cm \Rightarrow\hskip 0.6 cm
 \beta^\mu\beta_\mu=0\ .\eqno(4.4)$$

In the Dirac-Goto-Nambu model that is most familiar to present day
cosmologists, the eigenvector $\beta^\mu$ is indeterminate
and the energy density $U$ is the same (in relativistic units 
such as are used here, with the speed of light set to unity) as the 
corresponding (surface)  tension. However in general it is essential to 
distinguish the concept of energy density $U$ from the concept of the 
{\it tension} (as used in physics since the formulation of Hooke's law at the
time of Newton) from which term ``tensor'' is derived. The (surface) tension 
scalar, for which we shall use the traditional symbol $T$, is defineable
generically in a manner consistent with traditional usage, for a ``pure" 
p-brane (i.e. a (p+1)-dimensionally supported system) of the kind
considered here by decomposing the trace of the surface stress 
momentum energy density tensor in the form
$$\ov T{^\mu}{_\mu}=-U-{\rm p}T\ .\eqno(4.5)$$ 

Appart from the degenerate Dirac-Goto-Nambu case for which $U$ and $T$ are
actually equal, the simplest possibility is that of a {\it perfect} 
p-brane\cite{[4]},
meaning one whose surface stress momentum energy density tensor is 
spacially (thus p-dimensionally) {\it isotropic} so that it is will be
expressible, for a suitable choice of the normalisation of the eigenvector 
$\beta^\mu$ in (4.2) by 
$$\ov T{^{\mu\nu}}=\beta^\mu\beta^\nu-T\og^{\mu\nu}\ ,\eqno(4.6)$$
where the required normalisation is given by
$$U-T=\beta^\mu\beta_\mu\leq 0\ .\eqno(4.7)$$
This category includes the case of the Dirac-Goto-Nambu model (3.12), which
is obtained (with $T=m^{\rm p+1}$) by normalising the (in this case
indeterminate) eigenvector to zero, i.e setting $\beta^\mu=0$. A more
mundane example is provided by the familiar ``improper" case p=n--1 (where n
is the background dimension) of an ordinary (relativistic) perfect fluid
with pressure $P=-T$. Although the tension $T$ is negative in the ordinary
fluid case, it must be positive, as a condition for stability, for ``proper"
p-branes of lower dimension, p$<$n--1. Whereas the membrane models (with p=2
for n=4) that are appropriate for the description of nautical sails will not
in general have the isotropic form (4.6), a familiar everyday example of a
membrane that will automatically be ``perfect" in this sense is provided by
the case of an ordinary soap bubble whose boundary hypersurface will be
characterised by $0<T<<U$. The kind of application on which we shall
concentrate in the following sections is that of a {\it string}, as given by
p=1, for which the ``perfection" property (4.6) will always hold: isotropy
cannot fail to apply in this 1-brane case because only a single space
dimension is involved. 

Before specialising to the (automatically perfect) string case, it is to be
remarked that except when the $\beta^\mu$ is indeterminate 
(as in the Dirac-Goto-Nambu case), or null (as can occur in special subspaces
where the current becomes null in a string model of the 
kind\cite{[30]}\cite{[31]}\cite{[36]} appropriate for describing cosmic vortex 
defects of the ``superconducting" kind proposed by Witten\cite{[7]}) this 
eigenvector will determine a corresponding unit eigenvector,
 $$u^\mu=\big(U-T\big)^{-1/2}\beta^\mu\ , 
\hskip 1cm u^\mu u_\mu=-1 \ ,\eqno(4.8)$$
specifying a naturally preferred rest frame, in terms of which
(4.6) will be rewriteable in the standard form
$$\ov T{^{\mu\nu}}=\big(U-T\big)u^\mu u^\nu -T\og^{\mu\nu}\ .\eqno(4.9)$$
With respect to the frame so defined, it can be seen that the velocity
$c_{_{\rm E}}$ say of propagation of extrinsic perturbations
(what are commonly referred to as ``wiggles") of the perfect
brane worldsheet will be given, according to the characteristic equation
(3.10), by
$$c_{_{\rm E}}^{\ 2}={T\over U}\ .\eqno(4.10)$$
By substituting (4.8) in the (3.3), and using the expression (1.16) for the 
acceleration $\dot u^\mu$ of the preferred unit vector (4.8), the extrisic 
equation governing the free motion of the support surface of any (proper) 
perfect brane is reducible to the standard form\cite{[4]}\cite{[36]}
$$c_{_{\rm E}}^{\ 2} K^\mu=(1-c_{_{\rm E}}^2)\perp^{\!\mu}_{\,\nu}\dot u^\nu
\ .\eqno(4.11)$$
In the Dirac-Goto-Nambu case one has $U=T$ and hence $c_{_{\rm E}}^{\ 2}=1$,
(i.e. the ``wiggle" speed is that of light) so that the right hand side of
(4.11)  will drop out, leaving the harmonicity condition that is expressed
by the vanishing of the curvature vector given by (1.19).
At the opposite extreme from this  relativistic limit case, one has
applications to such mundane examples as that of an ordinary soap bubble
membrane or violin string for which one has $c_{_{\rm E}}^{\ 2}<<1$.

Except in the Dirac-Goto-Nambu limit case, the extrinsic equation of
motion (4.12) will need to be supplemented by the dynamic equations
governing the evolution of the internal fields, and in particular of
$T$, $U$, and $u^\mu$, on the brane surface. The simplest non trivial 
possibility is the case of a perfect brane model that is ``barotropic",
meaning that its tension $T$ (or equivalently, in the case of a fluid,
its pressure $P$) is a function only of the energy density $U$ in the
preferred rest frame. In this barotropic or ``perfectly elastic" 
case, the complete set of dynamical
equations governing the internal evolution of the perfect brane
will be provided just by the tangential projection of the force
balance law (2.24) which, in the free case to which the discussion of
the present section is restricted, has the simple form
$$\og^\lambda_{\ \nu}\ov\nabl_\mu\ov T{^{\mu\nu}}=0\ .\eqno(4.12)$$

In the barotropic case (which includes the kind of cosmic string
models\cite {[30]}\cite{[31]}\cite{[36]} appropriate for describing the 
Witten type superconducting vaccuum vortices\cite{[7]}
 whose investigation provided the original motivation for
developing the kind of analysis presented here) experience of the
``improper" case of an (n--1)-brane, i.e.  the familiar example of an
ordinary barotropic perfect fluid, suggests the convenience of introducing
an idealised particle {\it number density}, $\nu$ say, and a corresponding
(relativistic) {\it chemical potential}, or effective mass per idealised
particle, $\mu$ say, that are specified modulo an arbitrary normalisation
factor by the equation of state according to a prescription of the form 
$${\rm ln\ }\nu=\int{dU\over U-T} \ ,\hskip 1 cm {\rm ln\ }\mu
=\int{dT\over T-U}  , \eqno(4.13)$$
which fixes them modulo a pair of constants of integration which are
not allowed to remain independent but are related by the imposition of
the restraint condition
$$\mu\nu=U-T \ . \eqno(4.14)$$
The quantities introduced in this way can be used for defining a surface 
particle current density vector $\nu^\rho$ and a corresponding
dynamically conjugate energy - momentum (per particle) covector 
$\mu_\rho$ that are given by
$$\nu^\rho=\nu u^\rho\ ,\hskip 1 cm \mu_\rho=\mu u_\rho\ ,\eqno(4.15)$$
in terms of which the generic expression (4.9) of the surface stress momentum
energy density tensor of a perfect brane can be rewritten in the form
$$\ov T{^\rho}_{\!\sigma}=\nu^\rho\mu_\sigma-T\og^\rho_{\ \sigma}
\ ,\eqno(4.16)$$
whose divergence can be seen to be given identically by 
$$\ov \nabl_\rho\ov T{^\rho}_{\!\sigma}=\mu_\sigma\ov\nabl_\rho\nu^\rho
+2\nu^\rho\ov\nabl_{[\rho}\mu_{\sigma]}-T K_\sigma \ .\eqno(4.17)$$
In the free case for which this divergence vanishes, projection
orthogonal to the worldsheet gives back the extrinsic dynamical equation 
(4.11), while by contraction with $\nu^\rho$ one obtains the simple 
surface current conservation law
$$\ov\nabl_\rho\nu^\rho=0\ \ . \eqno(4.18)$$
The remainder of the system of equations of motion of the perfect brane can
hence be obtained from (4.12), as a surface generalised version
of the standard perfect fluid momentum transport law\cite{[51]}, in the form
$$\og^\rho_{\ \nu} u^\mu \ov\nabl_{[\mu}\mu_{\rho]}=0\ ,\eqno(4.19)$$
which is the integrability condition for the existence of a surface
potential function, $\varphi$ say, such that $\mu_\rho=\ov\nabl_\rho
\varphi$. 

Appart from the extrinsic perturbations of the world sheet location itself,
which propagate with the ``wiggle" speed $c_{_{\rm E}}$ (relative to the
frame deterined by $u^\mu$) as already discused, the only other kind of
perturbation mode that can occur in a barytropic string are longitudinal
modes specified by the variation of $U$ or equivalently of $T$ within the
world sheet. As in the ``improper" special case of an ordinary perfect
fluid, for which they are interpretable just as ordinary sound waves, such
longitudinal ``woggle" perturbations can easily be seen 
from (4.16) and (4.17) to have a relative propagation velocity 
$c_{_{\rm L}}$ say that is given\cite{[4]}\cite{[36]} by
$$   c_{_{\rm L}}^{\ 2}={\nu d\mu\over\mu d\nu}=-{dT\over dU}\ . 
 \eqno(4.20)$$
In order for a barotropic model to be well behaved, $c_{_{\rm E}}^{\ 2}$ and
$c_{_{\rm L}}^{\ 2}$ must of course both be positive, in order for
the velocities to be real (as a condition for local stability), and
they must also both be less than unity in order to avoid superluminal
propagation (as a condition for local causality).
An important qualitative question that arises for any {\it proper} 
barotropic p-brane, i.e. one with 1$\leq$ p$<$n--1 (excluding the trivial
case p=0 of a point particle, and at the other extreme the 
``improper" case p=n--1 of a
fluid) is whether the extrinsic propagation speed is subsonic, meaning
that ``wiggles" go slower than ``woggles", i.e
$c_{_{\rm E}} < c_{_{\rm L}}$, transonic, meaning
that the speeds are the same, $c_{_{\rm E}} =c_{_{\rm L}}$
or supersonic, meaning that ``wiggles" go faster than ``woggles", 
$c_{_{\rm E}} >c_{_{\rm L}}$.  Much of the earliest 
work\cite{[52]}\cite{[53]}\cite{[54]}\cite{[55]}\cite{[56]}\cite{[57]}\cite{[58]}
on the Witten type superconducting vaccuum vortex phenomena was implicitly
based on use of a string model of subsonic type, but a more careful analysis
of the internal structure of the Witten vortex by 
Peter\cite{[64]}\cite{[65]}\cite{[66]} has since
shown that models of supersonic type are more appropriate. On the other
hand such everyday examples as that of an ordinary violin string are
subsonic, a condition that is sufficient\cite{[67]} though not 
necessary\cite{[68]} for stability of the corresponding circular 
centrifugally supported loop 
configurations\cite{[69]}\cite{[70]}\cite{[71]}\cite{[72]}
that will be discussed later on.
 
A particularly important special case is that of a barytropic p-brane
of {\it permanently transonic} type, meaning one characterised by an equation of 
state of the non dispersive constant product form
$$UT=m^{\rm 2p+2} \ ,\eqno(4.21)$$
where $m$ is a constant having the dimensions of a mass, for which
application of the formulae (4.10) and (4.18) gives $c_{_{\rm E}}^{\
2}=c_{_{\rm L}}^{\ 2}$ not just for some critical transition value of $U$ or
$T$ (as can occur for other kinds of equation of state) but for all values.
In the string case, p=1, it can easily be demonstrated\cite{[4]} that this
non dispersive permanently transonic model is represents the outcome of a
(rather artificial) dimensional reduction process first suggested by
Nielsen\cite{[73]}\cite{[74]}\cite{[75]}\cite{[76]}. As will be described
below, this model can be shown\cite{[77]} to be governed by equations of
motion that are explicitly integrable in a flat empty background, and it can
be used\cite{[77]}\cite{[78]} to provide what (contrary to a misleading
claim\cite{[79]} that has been made) can be a highly accurate  description
of the macroscopic effect of ``wiggles" in an underlying cosmic string of
the simple Goto-Nambu type.

\section{Essentials of Elastic String Dynamics}

Between the hypersurface supported case of a membrane and the curve
supported case of a point particle, the only intermediate kind of brane that
can exist in 4-dimensions is that of 1-brane, i.e. a string model, which
(for any background dimension $n$) will have a first fundamental tensor that
is expressible as the square of the antisymmetric unit surface element
tensor ${\cal E}^{\mu\nu}=-{\cal E}^{\nu\mu}$ given by (1.28): 
$$\og^\mu_{\ \nu}={\cal E}^\mu_{\ \rho}{\cal E}^\rho_{\ \nu} \ 
. \eqno(5.1)$$
In the case of string (to which the remainder of this article will be
devoted) the symmetric surface stress momentum energy tensor
$\ov T{^{\mu\nu}}$ that is well defined for any ``pure" (i.e. unpolarised)
model will not only be expressible in the generic form (4.6) that expresses
the (in the string case trivial) property of spacial anisotropy with
respect to the preferred timelike or null eigenvector $\beta^\mu$, but
it will also be expressible generically in {\it bicharacteristic
form} as
$$\ov T{^{\mu\nu}}=\beta_{_+}^{[\mu}\beta_{_-}^{\,\nu]}\ ,\eqno(5.2)$$
in terms of a pair of timelike or null tangent vectors $\beta_{_\pm}^{\,\mu}$
that can be seen  to be extrinsic bicharacteric vectors, meaning that
they lie respectively allong the two (``right moving" and ``left moving")
directions of propagation of extrinsic perturbations,
and are thus respectively orthogonal to the covectors $\chi_\mu$
given by the extrinsic characteristic equation (3.10) whose solutions
can be seen from (5.2) to be given by $\beta_{_+}^{\,\mu}\chi_\mu=0$
or $\beta_{_-}^{\,\mu}\chi_\mu=0$. The expression (5.2) does not
completely determine the bicharacteristic vectors $\beta_{_\pm}^{\,\mu}$,
but leaves open the possibility of a reciprocal multiplicative
rescaling of their magnitudes, subject to the condition that their
scalar product remains invariant, its value being obtainable by comparison 
of (5.2) with (4.6) as
$$\beta_{_+\mu}\beta_{_-}^{\,\mu}=\ov T{^\nu}_{\!\nu}=-\big (U+T\big)\ .
\eqno(5.3)$$
The geometric mean of the magnitudes of the bicharacteristic vectors
remains similarly invariant, with value given by the magnitude of the 
preferred eigenvector $\beta^\mu$ as given by (4.7), i.e.
$$\big(\beta_{_+\mu}\beta_{_+}{^{\!\mu}}\big)\big(\beta_{_-\nu}
\beta_{_-}{^{\!\nu}}\big)=\big(\beta_\mu\beta^\mu\big)^2=(U-T)^2\ 
.\eqno(5.4)$$

To obtain the equations of motion of the string we need to use the
force balance equation (2.25), which just takes the form
$$\ov\nabl_\mu\ov T{^{\mu\nu}}=\ov f{^\nu}\ ,\eqno(5.5)$$ for a string that
is isolated, where $\ov f{^\mu}$ (which simply vanishes in the free case) is
the force exerted by any background fields, such as the electromagnetic
field and ambient (Higgs or other) fluid that are allowed for in the
explicit expression (2.26). (The only difference if the string were not
isolated, i.e. if it belonged to the boundary of one or more attached
membranes, is that an additional contact force contribution
$\check f{^\mu}$ of the form (2.25) would also be needed as well as the
background contribution $\ov f{^{\mu}}$.) Using the Weingarten integrability
condition (1.13) which is equivalent in the string case to the the projected
Lie commutativity condition
$$\perp^{\!\rho}_{\,\nu}\big(\beta_{_+}{^{\!\mu}}\ov\nabl_\mu
\beta_{_-}{^{\!\nu}}-\beta_{_-}{^{\!\mu}}\ov\nabl_\mu
\beta_{_+}{^{\!\nu}}\big)=0 \ ,\eqno(5.6)$$
it can be seen from (5.2) that the extrinsic equations of motion governing
the evolution of the string worldsheet will therefore\cite{[36]} be 
expressible in characteristic form as 
$$\perp^{\!\rho}_{\,\nu}\beta_{_\pm}{^{\!\mu}}\ov\nabl_\mu
\beta_{_\mp}{^{\!\nu}}=\perp^{\!\rho}_{\,\mu}\ov f{^\mu}\ ,\eqno(5.7)$$
for either choice of the sign $\pm$, one version being obtainable
from the other and vice versa by (5.6).

Except where one of the bicharacteristic vectors $\beta_{_\pm}{^{\!\mu}}$ is
null, which by (5.5) will only occur where $U=T$ (as is everywhere the case
for the Goto-Nambu model, for which both bicharacteristic vectors are null
everywhere) it will be possible to fix their normalisation by requiring that
they have the same magnitude, $\beta_{_\pm\mu}\beta_{_\pm}{^{\,\mu}}
=-(T-U)$ and hence to define a corresponding pair of {\it unit
bicharacteristic vectors} 
$$u_{_\pm}{^{\,\mu}}=\big (U-T\big)^{-1/2}\beta_{_\pm}{^{\,\mu}}
\ ,\hskip 1 cm     u_{_\pm\mu}u_{_\pm}{^{\,\mu}}=-1\ ,\eqno(5.8)$$
in terms of which the unit timelike eigenvector $u^\mu$ introduced in (4.8),
together with an orthogonal spacelike unit eigenvalue $v^\mu$, will be given 
by
$$u^\mu={\sqrt{1-c_{_{\rm E}}{^2}}\over 2}\big(u_{_+}{^{\,\mu}} +
u_{_-}{^{\,\mu}}\big)\ ,\hskip 1 cm v^\mu={\sqrt{1-c_{_{\rm E}}{^2}}
\over 2 c_{_{\rm E}}}\big(u_{_+}{^{\,\mu}} -u_{_-}{^{\,\mu}}\big)\ .
\eqno(5.9)$$

Such mutually orthogonal unit eigenvectors constitute a preferred diad, in
terms of which the antisymmetric unit surface element ${\cal E}^{\mu\nu}$
and the (first) fundamental tensor $\eta^{\mu\nu}$ of the string 
worldsheet will be given by
$${\cal E}^{\mu\nu}=u^\mu v^\nu-v^\mu u^\nu\ ,\hskip 1 cm
\og^{\mu\nu}=-u^\mu u^\nu+v^\mu v^\nu \ ,\eqno(5.10)$$
and in terms of which the standard form (4.9) for the surface stress 
momentum energy density tensor will reduce to the simple form
$$\ov T{^{\mu\nu}}=U u^\mu u^\nu - T v^\mu v^\nu \ .\eqno(5.11)$$
Proceeding directly from this last form, the extrinsic dynamical
equation (5.6), i.e. the surface orthogonal projection of (5.5),
can be rewritten as
$$\perp^{\!\mu}_{\,\nu}\big(U\dot u^\nu-T v^{\prime\nu}\big)=\perp^{\!\mu}
_{\,\nu}\ov f{^\nu}\ ,\eqno(5.13)$$
using the notation
$$\dot u^\mu=u^\nu\nabl_\nu u^\nu\ ,
\hskip 1 cm v^{\prime\mu}=v^\nu\nabl_\nu v^\nu \ .\eqno(5.14)$$
in terms of which the worldsheet curvature vector is expressible as 
$$K^\mu=\perp^{\!\mu}_{\,\nu}(v^{\prime\nu}-\dot u^\nu) \ .\eqno(5.15)$$ 

These last formulae illustrate a special feature distinguishing string models
from point particle models on one hand and from higher dimensional brane 
models on the other, namely the dual symmetry\cite{[30]} that exists at a 
formal level between the timelike eigenvector $u^\mu$  and the associated 
eigenvalue $U$ on one hand, and the the dynamically conjugate quantities
that are the spacelike eigenvector and $v^\mu$ and the associated eigenvalue
$T$. This feature is particularly informative in the barotropic or
``perfectly elastic" case
discussed at the end of the previous section, i.e. in the case when
the internal dynamics is controlled just by an equation of state giving
the energy density $U$ as a function of the tension $T$ only, since in
 there will be a corresponding duality relation between the
number density $\nu$ and the associated chemical potential $\mu$ as 
introduced in (4.13), which is equivalent to the mutually dual
pair of differential defining relations
$$dU=\mu d\nu\ , \hskip 1 cm dT=-\nu d\mu \ .\eqno(5.16)$$
Defining the surface duals of the current vector $\nu^\rho$ and momentum 
covector $\mu_\rho$ given by (4.15) as
$$\star\nu_\rho={\cal E}_{\rho\sigma}\nu^\sigma=\nu v_\rho\ ,\hskip 1 cm
\star\mu^\rho={\cal E}^{\rho\sigma}\mu_\sigma=\mu v^\rho\ ,\eqno(5.17)$$
one can use (4.17) to obtain the force balance equation for a barotropic
string model in the self dual form
$$\mu_\rho\ov\nabl_\sigma\nu^\sigma+\star\nu_\rho\nabl_\sigma\star\!\mu^\sigma
+\perp^{\!\sigma}_{\,\rho}\big(U\dot u^\sigma-T v^{\prime\sigma}\big)
=\ov f_\rho\ , \eqno(5.18)$$
whose orthogonally projected part gives back the self dual extrinsic 
force balance equation (5.13), while its tangentially projected part gives
a mutually dual pair of surface current source equations of the form
$$(U-T)\ov\nabl_\sigma\nu^\sigma=-\nu^\rho\ov f_\rho\ ,\hskip 1 cm
(U-T)\ov\nabl_\sigma\star\!\mu^\sigma=\star \mu^\rho\ov f_\rho \ .
\eqno(5.19)$$
It can be seen from this that, like the surface number current density
$\nu^\sigma$, the dual current $\star\mu^\sigma$ will also be conserved
in a free barotropic string, i.e. when the external force $\ov f_\rho$ is 
absent.

It is useful\cite{[30]} to introduce a dimensionless state function called 
the ``characteristic potential" $\vartheta$ say that is constructed in such 
a way as to be self dual according to the differential relation
$$d\vartheta=\sqrt{d\mu\, d\nu\over \mu\nu} \ .\eqno(5.20)$$ 
This function is convenient for the purpose of writing the internal
equations of motion obtained by tangential projection of (5.5) in a
characteristic form that will be the analogue of the characteristic version
(5.7) of the extrinsic equations of motion obtained by the orthogonal
projection of (5.5). The longitudinal bicharacteristic unit vectors
$\ell_{_\pm}{^{\!\mu}}$ say (the analogues for the internal sound type waves
of the extrinsic bicharacteristic unit vectors $u_{_\pm}{^{\!\mu}}$
introduced above) are defineable by
$$ \ell_{_\pm}{^{\!\mu}}=\big(1-c_{_{\rm L}}{^2}\big)^{-1/2}\big(u^\mu
\pm c_{_{\rm L}}v^\mu\big) \ ,\hskip 1 cm  \ell_{_\pm\mu} 
\ell_{_\pm}{^{\!\mu}}=-1\ , \eqno(5.21)$$
where  $c_{_{\rm L}}$ is the longitudinal characteristic speed as given
by (4.20). In terms of these,
  the tangentially projected part of the force balance equation 
(5.5) is convertible\cite{[30]} to the form of a pair of scalar  
equations having the characteristic form
$$(U-T)\ell_{_\pm}{^{\!\rho}}\big(\ov\nabl_\rho\vartheta \mp 
v_\nu\ov\nabl_\rho u^\nu\big)=\pm{\cal E}_{\mu\nu}\,\ell_{_\mp}{^{\!\mu}}\,
\ov f{^\nu}\ , \eqno(5.22)$$
from which it is directly apparent that, as asserted in the previous section,
the state function $c_{_{\rm L}}$, as given by (4.20), is indeed the 
characteristic speed of longitudinal ``woggle" propagation relative to the 
preferred frame specified by the timelike eigenvector $u^\mu$.  When 
rewritten in terms of the unit bicharacteristic vectors $u_{_\pm}{^{\!\mu}}$ 
that are the extrinsic analogues of $\ell_{_\pm}{^{\!\mu}}$, the 
corresponding extrinsic equations (5.7) take the form
$$(U-T)\perp^{\!\mu}_{\,\nu}u_{_\pm}{^{\!\rho}}\ov\nabl_\rho\,
u_{_\mp}{^{\!\nu}}=\perp^{\!\mu}_{\,\nu}\ov f{^\nu}\ ,\eqno(5.23)$$
(It is to be noted that the corresponding equation (32) at the end of the
original discussion\cite{[30]} of the characteristic forms of the equations
of string motion contains a transcription error whereby a proportionality
factor that should have been $\sqrt{(U-T)T}$ was replaced by its 
non-relativistic -- i.e. low tension -- limit, namely $\sqrt{UT}$.)

What essentially distinguishes different physical kinds of ``perfectly 
elastic", i.e. barotropic, string model from one  another is the form of 
the equation of state, which depends on the microsropic internal details of 
the vacuum defect (or other structure) that the string model is supposed to
represent. In a cosmological context, the most important special case,
which will be discussed in detail later on, is given by the constant
product formula (4.21), but for terrestrial applications the most
useful kind of equation of state for general purposes is the one named
after Robert Hooke, who is recogniseably the principal founder of string
theory as a branch of physics in the modern sense of the word.
(As for superstring theory, it remains to be seen whether it will become
established as a branch of physics or just as a topic of mathematics.)
Hooke's famous law to the effect that the tension, $T$ is proportional 
to the extension, i.e. the change in $1/\nu$ in the notation used 
here, provides a very good approximation in the low tension limit that 
is exemplified by applications such as that of an ordinary violin string.
This law is expressible within the present scheme as
$$U(Y+T)=U_0 Y+{_1\over^2} T^2\ ,\eqno(5.24)$$
where $U_0$ is a constant interpretable as the rest frame energy in the
fully relaxed (zero tension) limit, and $Y$ is another constant 
interpretable as what is known (after another pioneer worker on the 
subject) as Young's modulus.

A more general category of non-barotropic string models is of course
required, not only in such terrestrial applications as (above ground)
electric power transmission cables (the subject of my first research project
as an undergraduate working working for commercial industry during a
vacation),  but also for many cosmologically relevant cases that might be
envisaged (such as that of a ``warm superconducting string"\cite{[80]}) in
which several independent currents are present. Nevertheless most of the
cosmological string dynamical studies that have been carried out so far have
been restricted (not just for simplicity, but also as a very good
approximation in wide range of circumstances) to models of the the
``perfectly elastic" barotropic type, and more specifically to the category
(including the Hookean and Goto Nambu examples as opposite extreme special
cases) of string models describable by a Lagrangian of the form
$$\ov\L= \Lambda+\ov \j{^\mu}A_\mu+\ov\w{^{\mu\nu}}B_{\mu\nu}\ 
,\eqno(5.25)$$
in which, in order to satisfy the vorticity flux conservation requirement
(as derived in Section 2 from the condition of invariance with respect to 
gauge changes in the background Kalb-Ramond field $B_{\mu\nu}$) one
must have
$$\ov\w{^{\mu\nu}}=\kappa{\cal E}^{\mu\nu} \ ,\eqno(5.26)$$
for some fixed (quantised) circulation value $\kappa$ say, while another
fixed coupling constant, $e$ say, specifies the electromagnetic current
density, in the form
$$\ov \j{^{\mu}}=e c^{\mu}\ ,\eqno(5.27) $$
in terms of the only independent field variable on the world sheet, which
can be taken to be a stream function $\psi$ which enters only via
the corresponding identically conserved number density current
$$c^\mu={\cal E}^{\mu\nu}p_\nu \ , \hskip 1 cm  
p_\nu=\ov\nabl_\nu\psi\ ,\eqno(5.28)$$
whose squared magnitude,
$$\c=c_\mu c^\mu= -p^\mu p_\nu= - h^{ij}\psi_i\psi_j\ ,\eqno(5.29)$$
is the only argument in a (generically non-linear) {\it master function},
$$\Lambda=\Lambda\{ \c \}\ .\eqno(5.30)$$ 
Previous studies of cosmic strings have mostly been restricted to cases
involving allowance for a Kalb-Ramond coupling  (which is relevant for
vortex defects of ``global type" \cite{[26]}\cite{[81]}\cite{[82]} in the
absence of an internal current, or alternatively to cases (such as are
exemplified\cite{[30]}\cite{[31]} by Witten's local superconducting string
mechanism\cite{[7]}) involving allowance for an internal current in the
absence of Kalb-Ramond coupling. However there is no particular difficulty
in enviseaging the presence of both kinds of coupling simultaneously as is
done here. A new  result that is obtained thereby (in Section 6) is that for
stationary and other symmetric configurations, the effect of the  coupling
to tht Kalb Ramond background is expressible as that of a fictitious extra
electromagnetic field contribution.

\def\fff{\baselineskip=\normalbaselineskip}
%$ \vbox {\offinterlineskip{\hbox {$^{(p)}$} \hbox{T} }} $.

\def\spose#1{\hbox to 0pt{#1\hss}} 
\def\Libra{\spose {--} {\cal L}}
\def\Diam{\spose {\raise 0.3pt\hbox{+}} {\diamondsuit}  }

\def\eqdef{\fff\ \vbox{\hbox{$_{_{\rm def}}$} \hbox{$=$} }\ \thf }
\def\ov{\overline}
\def\oov{\offinterlineskip\overline}
\def\sta{{^\star}\!}
\def\oc{\check}
\def\ot{\widetilde}
\def\w{{ W}}            \def\j{{ J}}      \def\c{{\cal \chi}}
\def\dL{d_{_{\rm L}}}

\def\uth{{^{\,_{(3)}}}\!}  \def\utw{{^{\,_{(2)}}}\!}
\def\ud{{^{\,_{(\rm d)}}}\!}  \def\udi{{^{\,_{(\rm d-1)}}}\!}
\def\up{{^{\,_{(\rm p)}}}\!}  \def\udp{{^{\,_{(\rm d)}}}\!}

\def\og{\eta}

\def\nabl{\nabla\!} \def\onab{\ov\nabl}

\def\vv{_{\ \vert}}     \def\osb{\offinterlineskip\hbox}
\def\af{\fff\vbox{\osb{$_{_{\vv}}$} \hbox{$\ov {\ f} $} }\thf }
\def\ag{\perp\!}

\def\P{{\cal P}} \def\R{{\cal R}} \def\W{{\cal W}} \def\E{{\cal E}} 
\def\I{{\cal I}} \def\L{{\cal L}}
\def\A{{_A}} \def\B{{_B}} \def\X{{_X}} \def\Y{{_Y}}

\def\lag{{\cal L}}
\def\aag{{m^{\rm d}}} \def\bag{{\rm b}}   \def\cag{{\rm c}}

\def\pag{{\cal P}} \def\kag{{\mit\Sigma}}
\def\j{J}  \def\w{W}   \def\c{\chi}

It is the specific form of the master function that determines the relevant
equation of state -- or if its range of validity is sufficiently extensive,
the pair of equations of state, corresponding to the qualitatively different
regimes distinguished by the positivity or negativity of $\c$ -- by which
the dynamics of the model is governed. The derivative of the master function
provides a quantity 
$${\cal K}=2 {d\Lambda\over d\c}\ , \eqno(5.31)$$
in terms of which the surface stress momentum energy density tensor 
defined by (2.7) will be given by
$$\ov T{^{\mu\nu}}={\cal K} p^\mu p^\nu+\Lambda\og^{\mu\nu}\ .\eqno(5.32)$$
This only depends on the internal field gradient and
on the imbedding and the background geometry: it involves neither
the background electromagnetic and Kalb Ramond fields nor their respective
coupling constants $e$ and $\kappa$.  The role of the latter is just to
determine the background force density, which will be given, according to the
general formula (2.26), by
$$\ov f_\rho=e F_{\rho\mu} c^\mu+ {\kappa\over 2}N_{\rho\mu\nu}{\cal E}
^{\mu\nu} \ .\eqno(5.33)$$
Unlike the force density, the stress momentum energy density tensor, as 
given by (5.32), is formally invariant under a duality 
transformation\cite{[4]}\cite{[30]}
whereby the variational momentum covector $p_\mu$, the current $c^\mu$, 
and the master function, $\Lambda$ itself, are interchanged with their 
duals, as denoted by a tilde, which are definable by
$$\tilde c^\mu={\cal K} p^\mu \ ,\hskip  1 cm \tilde p_\mu={\cal K} c_\mu
\ , \hskip 1 cm \tilde\c=\tilde c_\mu\tilde c^\mu=-\tilde p^\nu
\tilde p_\nu \ ,\eqno(5.34)$$
and by the relation
$$\c{\cal K}=\Lambda-\tilde\Lambda=-\tilde\c \tilde{\cal K}\ 
,\eqno(5.35)$$
with
$$\tilde{\cal K}=2{d\tilde\Lambda\over d\tilde\c}={\cal K}^{-1}\ ,\hskip
1 cm \tilde \c=- {\cal K}^2\c \ .\eqno(5.36)$$
In terms of these quantities the stress momentum energy density tensor can 
be rewritten in either of the equivalent mutually dual canonical forms
$$\tilde c^\mu p_\nu+\Lambda\og^\mu_{\ \nu} = \ov T{^\mu}{_{\!\nu}}
=c^\mu\tilde p_\nu+\tilde\Lambda\og^\mu_{\ \nu} \ . \eqno(5.37)$$
Substitution of these respective expressions in (5.5) then gives the force
 balance equation in the equivalent mutually dual forms
$$p_\rho\ov\nabl_\mu\tilde c^\mu + 2\tilde c^\mu\ov\nabl_{[\mu} p_{\rho]}
+\Lambda K_\rho= \ov f_\rho = \tilde p_\rho\ov\nabl_\mu c^\mu + 2 c^\mu
\ov\nabl_{[\mu}\tilde p_{\rho]}+\tilde \Lambda K_\rho \ .\eqno(5.38)$$
The independent internal equations of motion of the system 
are obtainable by contraction with the independent mutually orthogonal 
tangent vectors $p^\rho$ and $\tilde p^\rho$, the latter giving what is
just a kinematic identity, since, by construction according to the 
prescription given above, one automatically has
$$\ov \nabl_\mu c^\mu= 0 = {\cal E}^{\rho\mu}\nabl_\mu p_\rho\ ,
\hskip 1 cm \tilde p^\rho f_\rho= 0\ .\eqno(5.39)$$
The only part of the internal equations of motion that is properly dynamical
(from the point of view of the variation principle  with respect to
$\Lambda$) is thus just the part got by contraction with $p^\rho$,
which gives
$$-\c \ov\nabl_\mu\tilde c^\mu=p^\rho\ov f_\rho=\c{\cal E}^{\rho\mu}
\ov\nabl_{[\mu}\tilde p_{\rho]}\ ,\hskip 1 cm  p^\rho\ov f_\rho=e \c {\cal E} 
^{\mu\rho} F_{\mu\rho}\ . \eqno(5.40)$$
It is to be noted that unlike the electromagnetic force, the Joukowski force
(interpretable as a manifestation of the Magnus effect) arising from the
Kalb-Ramond coupling in (5.33) will always act orthogonally to the world 
sheet and thus will not affect the internal dynamics of the string. It is 
evident that the internal force balance equation (5.40) can be rewritten 
simply as
$${\cal E}^{\mu\rho}\big( 2\nabl_{[\mu}\tilde p_{\rho]}+e F_{\mu\rho}\big)=0\ ,
\eqno(5.41)$$
which can be construed as the (Poincar\'e type) integrability condition for
the local existence of a scalar $\varphi$ on the worldsheet in terms of
which the dual momentum will simply be given as the tangentially projected
gauge covariant derivative: 
$$\tilde p_\rho=\ov D_\rho\varphi\ ,\hskip 0.6 cm \ov D_\rho=
\og^\mu_{\ \rho}D_\mu\ , \hskip 0.6 cm D_\mu=\nabl_\mu-eA_\mu
 \ .\eqno(5.42)$$
In applications to Witten type superconducting strings, this field $\varphi$
will be interpretable\cite{[30]} as being proportional to -- and therefore 
with respect to suitably scaled units equal to -- the phase of an underlying
complex bosonic field. The scalar $\varphi$ can be used as the primary 
independent variable, instead of $\psi$, in an alternative (dynamically 
conjugate) variational formulation using a Lagrangian constructed from 
$\tilde\Lambda$ instead of $\Lambda$.

If its range of definition is sufficiently extended, the master function
$\Lambda$ will determine not just one but a pair of distinct equations of
state, one applying to the ``magnetic" regime where the current $c^\mu$ is
spacelike so that $\c$ is positive, and the other applying to the
``electric" regime where the current $c^\mu$ is timelike so that the $\c$ is
negative.  In either of these distinct regimes -- though not on the critical
intermediate ``null state" locus where the current is lightlike so that $\c$
vanishes -- the mutually dual {\it canonical} forms (5.37) of the stress
momentum energy density tensor are replaceable by the equivalent more
manifestly symmetric {\it standard} form
$$\ov T^{\mu\nu}={\Lambda\over\c}\, c^\mu c^\nu+{\tilde\Lambda
\over\tilde\c}\,\tilde c^\mu\tilde c^\nu\ .\eqno(5.43)$$
By comparison with the corresponding expression (5.11) it can thus be seen
that the physical interpretation -- as energy density $U$, tension $T$,
particle number density $\nu$ and effective mass per particle $\mu$ -- of
the state functions introduced in the variational function theat has just
been given will be given, in the ``magnetic" regime where $c^\mu$ is
spacelike (the first possibility to be considered in the early work on the
Witten mechanism\cite{[7]}), by 
$$\Lambda=- T\ ,\hskip 0.6 cm \tilde\Lambda = - U\ , \hskip 0.6 cm
\c=\mu^2 \ ,\hskip 0.6 cm \tilde\chi=-\nu^2\ ,
\hskip 0.6 cm {\cal K}={\nu\over\mu}\ ,\eqno(5.44)$$
with
$$c^\rho=\mu v^\rho\ ,\hskip 0.6 cm \tilde c^\rho=\nu u^\rho\ ,\hskip 0.6 
cm p_\rho =\mu u_\rho\ ,\hskip 0.6 cm \tilde p_\rho=\nu v_\rho \ 
.\eqno(5.45)$$
On the other hand in the ``electric" regime where $c^\mu$ is spacelike,
one obtains what in general will be a different equation of state given
parametrically by the dual set of relations 
$$\Lambda=- U\ ,\hskip 0.6 cm \tilde\Lambda = - T \hskip 0.6 cm
\c=-\nu^2 \ ,\hskip 0.6 cm \tilde\chi=\mu^2\ ,\hskip 0.6 cm
{\cal K}={\mu\over\nu} \ ,\eqno(5.46)$$
with
$$c^\rho=\nu u^\rho\ ,\hskip 0.6 cm \tilde c^\rho=\mu v^\rho\ ,\hskip 0.6 
cm p_\rho =\nu v_\rho\ ,\hskip 0.6 cm \tilde p_\rho=\mu u_\rho \ 
.\eqno(5.47)$$

In realistic conducting string models one can not expect the appropriate
master function $\Lambda\{\c\}$ to be given exactly by a simple 
analytic formula, but only by the output of a detailed numerical
computation of the internal structure of the underlying vacuum defect
\cite{[83]}\cite{[64]}\cite{[65]}\cite{[66]}. There is however an 
artificial mechanism (proceeding by a Kaluza Klein type projection of a
simple Goto-Nambu model in an extended background with an extra space
dimension) first suggested by Nielsen\cite{[73]} that can easily be
shown\cite{[4]} to provide a particularly elegant illustration of the above
formalism with a master function given  (for an appropriate choice of the
scaling of the stream function $\psi$) in terms of a constant mass parameter
$m$ by
$$\Lambda=\sqrt{m^4-m^2\c}\ ,\hskip 0.6 cm \tilde\Lambda=
\sqrt{m^4 - m^2\tilde\c} \ .\eqno(5.48)$$
This has the rather exceptional but by no means unique property of providing
an equation of state, for the relationship between $U$ and $T$, that has
the {\it same} form in the magnetic regime $\c>0>\tilde\c$ as in the
electric regime $\tilde\c>0>\c$. What is quite unique about this
equation of state is its non-dispersive -- i.e. permanently 
transonic -- ``constant product" form (4.21), which can be seen\cite{[77]} 
to be expressible parametrically in terms of a dimensionless self dual state
function $\vartheta$ of the kind introduced in (5.20) by
$$U=m^2\sqrt{1+{\nu^2\over m^2}}=m^2{\rm coth\,}\vartheta\ ,\hskip 1 cm
T=m^2\sqrt{1-{\mu^2\over m^2}}=m^2{\rm tanh\,}\vartheta \ ,\eqno(5.49)$$
which gives
$$c_{_{\rm E}}=c_{_{\rm L}}={\rm tanh\,}\vartheta \ ,\eqno(5.50)$$
so that by (4.21) and (4.20), the extrinsic (``wiggle") and longitudinal
 (``woggle") bicharacteristic vectors will coincide, having the form
$$\ell_{_\pm}{^{\!\mu}}=u_{_\pm}{^{\!\mu}} ={\rm cosh\,}\vartheta\, 
u^\mu \pm {\rm sinh\,}\vartheta\, v^\mu \ .\eqno(5.51)$$
The revelation\cite{[4]} of this unique transonicity property
invalidates the too hasty claim\cite{[74]} that Nielsen's elegant 
artifice\cite{[73]} effectively represents the outcome in the ``pure" 
string limit (in which effects of finite vortex thickness are neglected) of
the Witten mechanism\cite{[7]}. The latter requires a model\cite{[83]} of 
generically dispersive type\cite{[30]} that has been shown by the work of 
Peter\cite{[64]}\cite{[65]}\cite{[66]} to be characterised typically
by supersonicity, $c_{_{\rm E}}>c_{_{\rm L}}$. The elegant 
transonic model can nevertheless provide a crude but mathematically convenient
approximation to a realistic description of a superconducting string that 
should be a considerable improvement not only on the use\cite{[84]}
for this purpose of an unmodified Goto-Nambu model, but also on the
more commonly used description provided by the naively linearised
model\cite{[52]}\cite{[53]}\cite{[54]}\cite{[55]}\cite{[56]}\cite{[57]}\cite{[58]}
with equation of state given by constancy of the trace (5.3) which,
by (4.10) and (4.20), is evidently characterised by permanent subsonicity,
 $c_{_{\rm E}}<c_{_{\rm L}}=1$.

In terms of the bicharacteristic vectors (5.51), the intrinsic equations 
of motion (5.19) can be recombined\cite{[12]}\cite{[80]} as an 
equivalent pair of divergence  relations in the form
$$\ov\nabl_\rho\Big( (U-T)\ell_{_\pm}{^{\!\rho}}\Big)
=-\ell_{_\mp}{^{\!\rho}}\ov f_\rho \ ,\eqno(5.52)$$
which shows that the  ``left" and ``right" moving ``bicharacteristic
currents", $(U-T)\ell_{_+}{^{\!\rho}}$ and $(U-T)\ell_{_-}{^{\!\rho}}$ will
each be conserved separately in the free case, i.e. when $\ov f_\rho$
vanishes.  As an alternative presentation of the tangential force balance
equations for this permanently transonic model, a little algebra suffices to
show that the intrinsic equation of motion (5.22) can be rewritten for this
case, in a form more closely analogous to that of the corresponding
extrinsic equation of motion (5.23), as the pair of equations
$$(U-T)\og^{\mu}_{\ \nu}\,\ell_{_\pm}{^{\!\rho}}\ov\nabl_\rho \ell_{_\mp}
{^{\!\nu}}=\big(\og^{\mu}_{\ \nu}+ \ell_{_\mp}{^{\!\mu}} 
\ell_{_\mp\nu}\big) \ov f{^\nu}\ .\eqno(5.53)$$
This tangentially projected part of the dynamic equations can now be 
recombined\cite{[77]} with its orthogonally projected analogue (5.23),
so as to give the {\it complete} set of force balance equations for the
non dispersive permanently transcharacteristic string model (4.21) as 
the extremely useful pair of bicharacteristic propagation equations
$$(U-T)\ell_{_\pm}{^{\!\rho}}\ov\nabl_\rho \ell_{_\mp}{^{\!\mu}}=
\big(g^{\mu\nu} + \ell_{_\mp}{^{\!\mu}} \ell_{_\mp}{^{\!\nu}}\big)
\ov f{_\nu}\ .\eqno(5.54)$$

The advantage of the characteristic formulation (5.54) is that it takes a 
particularly simple form when expressed in terms of the corresponding
characteristic coordinates $\sigma^{_\pm}$ on the worldsheet as defined
by taking the ``right moving" coordinate $\sigma^{_+}$
to be constant allong ``left moving" characteristic curves, and taking the
``left moving" coordinate $\sigma^{_-}$ to be constant along ``right
moving" characteristic curves, with the convention that the correspondingly
parametrised bicharacteristic tangent vectors 
$\ell_{_\pm}^{\,\mu}\equiv{\partial x^\mu/\partial \sigma^{_\pm}}$ 
should be future directed. In the case of a free motion (i.e. when
the force term on the right of (5.54) vanishes) in a flat background,
this simplification can be used\cite{[77]} to obtain the complete solution
of the dynamical equations in a very simple explicit form. The way this 
works is that (5.54) reduces just to
$$  {\partial \ell_{_\pm}^{\,\nu}/\partial \sigma_{_\mp}}=0 \ ,\eqno(5.55)$$ 
whose general solution is given in terms of a pair of generating curves
$x_{_\pm}^{\,\mu}\{\sigma\}$ as a sum of single variable functions by the
ansatz
$$ x^\mu=x_+^{\,\mu}\{\sigma^{_+}\} +x_-^{\,\mu}\{\sigma^{_-}\}\ 
,\eqno(5.56)$$
which gives $\ell_{_\pm}^{\,\mu}=\dot x_{_\pm}^{\,\mu}$
using the dot here to denote the ordinary derivatives of the single variable
functions with respect to the corresponding characteristic variables. The
solution (5.56) generalises a result that is well known for more familiar
but degenerate Goto Nambu case, in which the tangents to the generating
curves are required to be null, $\ell_{_\pm}^{\,\mu}\ell_{_\pm\mu}=0$ (so
that with the usual normalisation their space projections lie on what is
known as the Kibble Turok sphere\cite{[85]}). The only restriction in the
non - degenerate case is that they should be non-spacelike and future
directed (the corresponding projections thus lying anywhere in the interior,
not just on the surface, of a Kibble Turok sphere) the unit normalisation
condition (5.21) being imposable as an option, not an obligation, by
choosing the parameter $\sigma$ to measure {\it proper} time allong each
separate generating curve.

This special property of being soluble by an ansatz of the same form (5.56)
as has long been familiar for the Goto Nambu case can immediately be used to
provide a new direct demonstration of the validity of the non-dispersive
constant product model (4.21) for describing the average motion of a
``wiggly" Goto Nambu string. The previously unpublished justification
presented here is needed because my original argument\cite{[77]} was merely
of a qualitative heuristic nature, while Vilenkin's mathematical
confirmation\cite{[78]} was based on indirect energetic considerations, and
has been called into question\cite{[79]} on the grounds that it did not
cover the most general class of wiggles that can be envisaged. It has
recently been shown by Martin\cite{[86]} that Vilenkin's method\cite{[79]}
(extending to the tension $T$ the concept of ``renormalisation" that had
previously been introduced for the energy density $U$ by Allen and
Shellard\cite{[87]}\cite{[88]}) can in fact be generalised
straightforewardly so as to confirm the validity of (4.21) for all kinds of
``wiggle" perturbations (subject only to the restriction that their
amplitudes should not be so large as to bring about a significant rate of
self intersection). Though based on manifestly muddled reasonning, the
purported contrary demonstration\cite{[79]} of higher order ``deviations"
from the constant product form  (4.21) for the effective equation of state
has left a residue of controversy and confusion. The need for an absolutely
clear refutation of such allegations has motivated the formulation, as a
more direct alternative to the energetic analysis developed by Vilenkin and
Martin, of the new derivation presented here.

The new justification for the use of the elastic string model characterised
by (4.21) consists simply of the observation that such a model implicitly
underlies the diamond lattice discretisation that, since its original
introduction by Smith and Vilenkin\cite{[89]}, has been commonly employed by
numerical simulators\cite{[90]} as a very convenient approximation scheme
-- of in principle unlimited accuracy -- for the representation of a
Goto-Nambu string worldsheet. As a way of replacing the exact continouous
description by a discrete representation such as is necessary for numerical
computation, the idea of the Smith Vilenkin method is simply to work with a
pair of discrete sets of sampling points  $x_{_\pm\rm r}^{\,\mu}=
x_{_\pm}^{\mu}\{\sigma_{\rm r}\}$ determined by a corresponding
discrete set of parameter values $\sigma_{\rm r}$ on
the generating curves of the exact representation (5.56).
This provides a ``diamond lattice" of sample points given (for integral
values of r and s) by
$$x^\nu_{\rm rs}=x_{_+\rm r}^{\,\mu}+x_{_-\rm s}^{\,\mu}\ ,\eqno(5.57) $$
that will automatically lie exactly on the ``wiggly"  Goto Nambu worldsheet
(5.56), which is thus represented to any desired accuracy by choosing a
sufficiently dense set of sampling parameter values $\sigma_{\rm r}$ on the
separate ``wiggly" null generating curves $x_{_\pm}^{\,\mu}\{\sigma\}$. The
new remark I wish to make here is simply that the chosen set of sample
points $x_{_\pm\rm r} ^{\,\mu}=x_{_\pm}^{\mu}\{\sigma_{\rm r}\}$ on the
separate ``wiggly" null generators can also be considered to be sample
points on a pair of {\it smoothed out}, and thus no longer null but {\it
timelike}, interpolating curves that, according to the result\cite{[77]}
demonstrated above, can be interpreted according to (5.56) as generating a
corresponding solution of the equations of motion for an elastic string
model of the kind governed by (4.21).  The not so ``wiggly" elastic string
worldsheet constructed by this smoothing operation will obviously be an even
better approximation to the exact ``wiggly" Goto Nambu worldheet than the
original Smith Vilenkin lattice representation, which itself could already
be made as accurate as desired by choosing a sufficiently high sampling
resolution. No matter how far it is extrapolated to the future, the smoothed
elastic string worldsheet generated according to (5.55) can never deviate
significantly from the underlying ``wiggly" Goto-Nambu worldsheet it is
designed to represent because the exact worldsheet and the smoothed
interpolation will always coincide precisely at each point of their shared
Smith Vilenkin lattice (5.57). This highly satisfactory feature of providing
a potentially unlimited accuracy could not be improved but would only be
spoiled by any ``deviation" from the originally proposed\cite{[77]} form
(4.21) for the effective equation of state.

After thus conclusively establishing that the permanently transonic
elastic string model characterised by the simple constant product equation
of state (4.21) (without any higher order corrections) provides
an optimum description of the effect of microscopic wiggles
in an underlying Goto Nambu model so long as self intersections 
remain unimportant (as was assumed in all the 
discussions\cite{[77]}\cite{[78]}\cite{[79]}\cite{[86]} cited above),
it remains to be emphasised that the neglect of such intersections will 
not be justified when the effective temperature\cite{[77]}\cite{[80]} 
of the wiggles is too high (as will presumably be the case\cite{[12]} 
during a transient period immediately following the string - forming 
phase transition). The result of such intersections will be the formation 
of microscopic loops, of which some will subsequently be reconnected, 
but of which a certain fraction will escape. Estimation of the 
dissipative cooling force density that would be needed to allow 
for such losses remains a problem for future work.

\section{Symmetric Configurations including Rings and their Cosmological
Implications}

Whenever the background space time metric is invariant under the action of a
(stationarity, axisymmetry, or other) continuous invariance group
generated by a solution $k^\mu$ Killing  equation (2.31), i.e. in the
notation of (2.20)
$$\vec{\ k\Libra} g_{\mu\nu}= 0\ ,\eqno(6.1)$$
then any string that is isolated (i.e. not part of the boundary of an 
attached membrane) will have a corresponding {\it momentum} current  
(interpretable, depending on the kind of symmetry involved, 
as representing a flux of energy, angular momentum, or whatever) given  by
$$\ov{P}{^\mu}=\ov T{^\mu}_{\!\nu} k^\nu\ .\eqno(6.2)$$
In accordance with (2.33), this will satisfy a source equation of the 
form
$$\ov\nabl_\mu\ov{P}{^\mu}= \ov f_\mu k^\mu\ , \eqno(6.3)$$
which means that the corresponding flux would be strictly conserved when
the string were not just isolated but {\it free}, i.e. if the
background force $\ov f_\mu$ were zero. When the string is subject to a
background force of the Lorentz-Joukowsky form (5.33) that arises from
background electromagnetic and Kalb-Ramond fields, then provided these
background fields are also invariant under the symmetry group action
generated by $k^\mu$, i.e. in the notation of (2.20)
$$\vec{\ k\Libra} A_{\mu}= 0\ ,\hskip 1 cm \vec{\ k\Libra} B_{\mu\nu}= 0
\ ,\eqno(6.4)$$
it can be seen to follow that, although the physically well defined surface
current $\ov {P}{^\mu}$ will no longer be conserved by itself, it still 
forms part of a gauge dependent generalisation, $\ov{\cal P}{^\mu}$ say, 
that is strictly conserved, 
$$\ov\nabl_\mu\ov{\cal P}{^\mu}=0\ ,\eqno(6.5)$$
and that is given, in terms of the gauge dependent generalisation
$$\ov{\cal T}{^\mu}_{\!\nu}=\ov T{^\mu}_{\!\nu}+\ov\j{^\mu}A_\mu+
\ov\w{^{\mu\rho}}B_{\nu\rho} \eqno(6.6)$$
of the surface stress momentum energy density tensor, by
 $$\ov{\cal P}{^\mu}=\ov{\cal T}{^\mu}_{\!\nu}k^\nu=\ov{P}{^\mu}+
\big(e c^\mu A_\nu +\kappa{\cal E}^{\mu\rho}B_{\nu\rho}\big)k^\mu
\ . \eqno(6.7)$$

The purpose of this section is to consider string configurations that share
the background symmetry under consideration. A configuration
that is symmetric in this sense will be characterised by the condition
$$\perp^{\!\mu}_{\,\nu}k^\nu=0\ ,\eqno(6.8)$$
meaning that the symmetry generator $k^\mu$ is tangential to the 
worldsheet, and the corresponding Lie invariance condition on
its surface stress momentum energy density tensor will have the form
$$k^\mu\nabl_\mu\ov T{^{\nu\rho}}=2 \ov T{^{\mu(\nu}}\nabl_\mu k^{\rho)}
\ .\eqno(6.9)$$
Under such conditions, as well as the ordinary momentum flux $\ov{P}{^\mu}$, 
what may be termed the {\it adjoint momentum} flux,
$$^\dagger\!\ov{ P}{^\mu}=k^\nu {^\dagger}\ov{T}{_\nu}{^\mu}\ ,
\hskip 1 cm ^\dagger\ov{T}{_\nu}{^\mu}={\cal E}_{\nu\rho}
{\cal E}^{\mu\sigma}\ov{ T}{^\rho}_{\!\sigma} \ ,\eqno(6.10)$$
will also obey an equation of the form (6.3), i.e.
$$^\dagger\ov\nabl_\mu\ov{P}{^\mu}= \ov f_\mu k^\mu\ , \eqno(6.11)$$
and  hence would also be conserved if the string were free.
In the presence of an electromagnetic or Kalb Ramond background, it
can be seen that, like the ordinary momentum flux, this adjoint momentum flux
has a gauge dependent extension,
$$^\dagger\!\ov{\cal P}{^\mu}=k^\nu {^\dagger}\ov{\cal T}{_\nu}{^\mu}\ ,
\hskip 1 cm ^\dagger\ov{\cal T}{_\nu}{^\mu}={\cal E}_{\nu\rho}
{\cal E}^{\mu\sigma}\ov{\cal T}{^\rho}_{\!\sigma} \ ,\eqno(6.12)$$
that will share with the generalised momentum flux
$\ov{\cal P}{^\mu}$ (that would be conserved even if the string did not
share the symmetry of the background) the property of obeying
a strict surface current conservation law, namely
$$\ov\nabl_\mu\, ^\dagger\!\ov{\cal P}{^\mu}=0\ .\eqno(6.13)$$

The group invariance conditions 
$$k^\nu\nabl_\nu p_\mu+p_\nu\nabl_\mu p^\nu=0\ ,\hskip 1 cm
k^\nu\nabl_\nu \tilde p_\mu+\tilde p_\nu\nabl_\mu p^\nu=0 \ ,\eqno(6.14)$$
that are the analogues of (6.9) for the separate mutually dual pair of
internal momentum covectors $\tilde p_\mu$ and $ p_\mu$
associated with the internal current within the string, can be rewritten,
with the aid of the corresponding electromagnetic background invariance 
condition (6.4), in the form
$$k^\rho{\cal E}^{\mu\nu}\big(\nabla_\nu\tilde p_\mu+{e\over 2}F_{\nu\rho}
\big)={\cal E}^{\rho\nu}\nabl_\nu\omega\ , \hskip 1 cm
k^\rho{\cal E}^{\mu\nu}\nabla_\nu p_\mu={\cal E}^{\rho\nu}\nabl_\nu\beta
  \ ,\eqno(6.15)$$
where $\omega$ and $\beta$ are are the Bernoulli type scalars given by
$$\omega=\tilde\beta +eA_\mu k^\mu\ ,\hskip 1 cm \tilde\beta=\tilde p_\mu 
k^\mu \ , \hskip 1 cm \beta =p_\mu k^\mu \ .\eqno(6.16)$$
If $k^\mu$ is timelike, the corresponding symmetry will be interpretable
as {\it stationarity}, while the more restrictive case\cite{[91]} in
which the string is actually {\it static} (in the sense that 
there is no transverse current  component relative to the background rest 
frame determined by $k^\mu$) will be given by the condition that the
second Bernouilli constant, $\beta$ should vanish.

It can be seen from (6.15) that the internal equations of motion (5.39)
and (5.41) are equivalent in this group invariant case simply to the
corresponding pair of Bernouilli type conservation laws to the effect
that $\omega$ and $\beta$ (but, unless the electromagnetic field is 
absent, not $\tilde\beta$) should both be {\it constant} over the worldsheet.
This observation allows the problem of solving the dynamical for a 
symmetric configuration of the kind of (``perfectly elastic" , i.e.
barocentric) string model under consideration to one of solving just
the extrinsic equations governing the location of the worldsheet. A recent 
investigation\cite{[92]} based on the systematic use of variational methods
in the restricted case for which the Kalb-Ramond coupling was absent
has drawn attention to the interest of extrapolating the Bernoulli constants
outside the supporting worldsheet as a pair of scalar fields defined over
the entire background spacetime by the uniformity conditions
$$\nabl_\mu\omega=0\ , \hskip 1 cm \nabl_\mu\beta=0\ ,\eqno(6.17)$$
and formulating the problem in terms a certain particular worldsheet tangent
vector that is defined by
$$X^\mu={\cal E}^{\mu\nu}{^\dagger}\ov{P}_\nu=k^\nu{\cal E}_{\nu\rho}
\ov T{^{\rho\mu}}\ .\eqno(6.18)$$
This vector will be expressible in terms of the variables introduced in
previous section as
$$X^\mu={\beta\Lambda\over\chi} c^\mu+{\tilde\beta\tilde\Lambda
\over\tilde\chi}\tilde c^\mu\ ,\eqno(6.19)$$
while the Killing vector itself will be expressible in analogous form by
$$k^\mu={\tilde\beta\over\Lambda-\tilde\Lambda}\,c^\mu+{\beta\over\tilde\Lambda
-\Lambda}\,\tilde c^\mu \ .\eqno(6.20)$$
The latter gives for the (real or pure imaginary) Killing vector amplitude,
$V$ say (which in the case of stationary symmetry, for which it is real,
will be interpretable as an effective gravitational potential field), an 
expression of the form
$$V^2=-k^\mu k_\mu={\beta^2\over\chi}+{\tilde\beta^2\over\tilde\chi}
\ ,\eqno(6.21)$$
that, in view of (6.16), and of the state functional relationship
between $\chi$ and $\tilde\chi$,  can be solved for any particular choice
 of the constant``tuning parameters" $\omega$ and $\beta$ to determine the
internal variables $\chi$ and $\tilde \chi$ as functions of  the 
(gravitational type) potential $V$ and the (electric type) potential
$A_\mu k^\mu$, and hence by implication as scalar fields over the entire 
background space, not just on the worldsheet where they were originally defined.
In a similar manner, it can be seen that the contraction of the worldsheet 
generating vector $X^\mu$ with the Killing vector field will be given 
simply by
$$k^\mu X_\mu=\beta\tilde\beta\ ,\eqno(6.22)$$
and that its (real or pure imaginary) amplitude, $X$ say, will given by an 
expression of the form
$$X^2=X^\mu X_\mu={\beta^2\Lambda^2\over \chi}+{\tilde\beta^2
\tilde\Lambda^2\over\tilde\chi} \, \eqno(6.23)$$
whereby it too is implicitly defined as a function of $V$ and $A_\mu 
k^\mu$, and hence as a well defined scalar field, not just on the worldsheet
but, by (6.16), also over the background as a whole, its gradient
being given by
$$X\nabl_\mu X=\Lambda\tilde\Lambda\, V\nabl_\mu V+e\tilde\beta(\Lambda
-\tilde\Lambda){\tilde\Lambda\over\tilde\chi} \nabla_\mu(A_\nu k^\nu)
\ .\eqno(6.24)$$
Previous experience\cite{[92]} with the case in which only the electromagnetic
but not the Magnus force contribution is present suggests the interest
of formulating the problem in terms of the propagation of the special
generating vector $X^\mu$, which (using the formulae (5.5) and (5.43)
 of the previous section) can be seen from (6.19) and (6.20) to be given by
$$X^\nu\nabl_\nu X_\mu=\Lambda\tilde\Lambda\, k^\nu\nabl_\nu k_\mu
+X^\nu{\cal E}_{\nu\rho}k^\rho \ov f_\mu \ ,\eqno(6.25)$$
where $\ov f^\mu$ is the background force as given by (5.33). It can now 
be seen that the two preceeding equations can be combined to give the 
equation of motion for the worldsheet generating vector $X^\mu$ in
the very elegant and convenient final form
$$X^\nu\nabl_\nu X_\mu-X\nabl_\mu X={\cal F}_{\mu\nu}
X^\nu\ ,\eqno(6.26)$$
in terms of a pseudo Maxwellian field given by
$${\cal F}_{\mu\nu}=e\beta F_{\mu\nu}+\kappa N_{\mu\nu\rho}k^\rho
=2\nabl_{[\nu}{\cal A}_{\rho]}\ ,\eqno(6.27)$$
where ${\cal A_\mu}$ is a gauge dependent pseudo-Maxwellian potential
covector given by
$${\cal A}_\mu=e\beta A_\mu+\kappa B_{\mu\nu}k^\nu\ .\eqno(6.28)$$
If $k^\mu$ is timelike so that the corresponding symmetry is interpretable
as {\it stationarity}, then the equation of motion (6.26) will be 
interpretable as the condition for the string to be in {\it equilibrium} 
with the given values of the constant ``tuning" parameters $\omega$ and 
$\beta$, (of which, as remarked above, the latter, $\beta$, will vanish
in the case of an equilibrium that is not just stationary but 
{\it static}\cite{[91]}). The new result here is that the Joukowski type
``lift" force (which was not allowed for in the previous analysis\cite{[92]})
due to the Magnus effect on the string, as it ``flies" (like an aerofoil) 
through the background medium represented by the current 3-form 
$N_{\mu\nu\rho}$, has just the same form as an extra Lorentz type 
electromagnetic (indeed in the stationary case purely magnetic) force 
contribution. 

When the genuine electromagnetic background coupling, and the similarly 
acting Kalb Ramond coupling  are both absent, then (as pointed out 
previously\cite{[92]}) the equation of motion (6.26) (that is the 
equilibrium condition for ``steady flight" in the stationary case) 
is just a simple  geodesic equation with  respect to, not the actual
background spacetime metric $g_{\mu\nu}$, but the conformally
modified metric $X^2 g_{\mu\nu}$, with the conformal factor $X^2$ determined
as a field over the background by (6.23) in conjunction with (6.17) and
(6.21). Even when the Lorentz and Joukowski force contributions are present,
the equation (6.26) governing the propagation of the world sheet generator
$X^\mu$ retains a particularly convenient Hamiltonian form, given by
$$ X^\mu={dx^\mu\over d\sigma}={\partial H\over\partial{\mit\Pi}_\mu}
\ ,\hskip 1 cm {d{\mit\Pi}_\mu\over d\sigma}=-{\partial H\over\partial x^\mu}
\ ,\eqno(6.29)$$
for the quadratic Hamiltonian function
$$H={_1\over^2}g^{\mu\nu}({\mit\Pi}_\mu-{\cal A}_\mu)
({\mit\Pi}_\nu-{\cal A}_\nu) -{_1\over^2}X^2 \ ,\eqno(6.30)$$
subject to a restraint fixing the (generically non-affine) parametrisation 
$\sigma$ of the trajectory by the condition that the numerical value of the
Hamiltonian (which will automatically be a constant of the motion)
should vanish, 
$$ H=0\ , \eqno(6.31)$$
together with a further momentum restraint, determining the (automatically 
conserved) relative transport rate ${\mit\Pi}_\mu k^\mu$ in accordance 
with the relation (6.22) by the condition
$${\mit\Pi}_\mu k^\mu=\omega\beta\ .\eqno(6.32)$$
The Hamiltonian momentum covector itself can be evaluated as
$${\mit\Pi}_\mu=X_\mu+{\cal A}_\mu={\cal E}_{\mu\nu}
{^\dagger}\ov{\cal P}^\nu\ ,\eqno(6.33)$$
or more explicitly, in 
terms of the original conserved generalised momentum flux $\ov{\cal P}
{^\mu}$ as  given by (6.7), and the gradients of the scalar (stream 
function and phase) potentials introduced in (5.28) and (5.42), as
$${\mit\Pi}_\mu= k^\nu{\cal E}_{\nu\rho}\ov{\cal T}
{^\rho}_{\mu}=\ov{\cal P}{^\nu}{\cal E}_{\nu\mu}+\omega\nabl_\mu\psi
+\beta\nabl_\mu\varphi\ .\eqno(6.34)$$
The advantage of a Hamiltonian formulation is that it allows the problem
to be dealt with by obtaining the momentum covector in the form
${\mit\Pi}_\mu=\nabl_\mu S$ from a solution of the corresponding Hamilton 
Jacobi equation, which in this case will take the form
$$g^{\mu\nu}\big(\nabl_\mu S-{\cal A}_\mu\big)\big(\nabl_\nu S-{\cal A}_\nu
\big)=X^2\ ,\eqno(6.35)$$
with $X^2$ given by (6.23) via (6.21), while the restraint (6.32) gives the
condition
$$k^\nu\nabl_\nu S=\beta\omega\ .\eqno(6.36)$$

Generalising results obtained previously\cite{[93]}\cite{[94]} for the Goto
Nambu limit case, it has recently been shown\cite{[92]} that for a string
model of the non-dispersive permanently trasonic type with the constant
product equation of state (4.21) that is governed by the Lagrangian (5.48)
(as obtained\cite{[4]} both from the Nielsen dimensional reduction mechanism
and also\cite{[77]}, as explained in Section 5,  from the more physically
realistic ``wiggly" string approximation) the stationary Hamilton Jacobi
equation is {\it exactly soluble by separation of variables} in a Kerr black
hole spacetime, not just of the ordinary asymptotically flat kind but even
of the generalised asymptotically De Sitter kind\cite{[95]}\cite{[96]}.
Except in the Schwarschild-De Sitter limit, where it could of course have
been predicted as a consequence of spherical symmetry, this separability
property still seems rather miraculous, reflecting a ``hidden symmetry" of
the Kerr background that is still by no means well understood. The newly
discovered separability property\cite{[92]} is not just an automatic
consequence of the simpler, though when first discovered already surprising,
property of separability for the ordinary geodesic equation\cite{[97]} but
depends on a more restrictive  requirement of the kind needed for the more
delicate separability property of the scalar wave equation\cite{[98]}. (It
is however more robust than the separability properties that have turned out
to hold for higher spin bosonic\cite{[99]} and
fermionic\cite{[100]}\cite{[101]}\cite{[102]} wave equations, and other
related systems\cite{[103]}\cite{[104]}.)

The most cosmologically important application of the formalism that has just
been presented is to the equilibrium of small {\it closed string loops} in
the mathematically relatively trivial case for which the background
gravitational, electromagnetic, and Kalb-Ramond fluid are negligible, so
that the trajectories generated by solutions of (6.26) will all just be {\it
straight lines} in a Minkowski background. Since in such a background the
Killing vector trajectories are also straight in the case of stationarity
(though not of course for axisymmetry) this might at first be perceived
implying that the worldsheet of a stationary string in an empty Minkowski
background would necessarily be flat. This conclusion would exclude the
possibility of closed loop equilibrium states in the absence of a background
field, and indeed cosmologists seem (albeit for other reasons) to have
entirely overlooked the possibility that such states might exist until the
comparitively recent publication of an epoch making paper by Davis and
Shellard\cite{[105]} provided the first counterexamples (the only previously
considered equilibrium states\cite{[106]}\cite{[54]}\cite{[55]}\cite{[56]}
having been based on a magnetic support mechanism that was was finally
judged to be too feeble to be effective except\cite{[109]} as a minor
correction).

The loophole in the deduction that if the trajectories generated by $X^\mu$
and by $k^\mu$ are both straight then the worldsheet must be flat is that it
is implicitly based on the assumption that the two kinds of trajectories
cross each other transversly. However there will be no restriction on the
curvature in the transverse direction in the critical case for which the two
kinds of trajectory coincide, i.e. for which $X^\mu$ and $k^\mu$ are
parallel. The condition for criticality in this sense is expressible as
$$X^\mu{\cal E}_{\mu\nu}k^\nu=0 \ ,\eqno(6.37)$$
which can be seen from the original definition (6.18) to be interpretable as
meaning that the tangent covector $\chi_\mu={\cal E}_{\mu\nu}k^\nu$
satisfies the extrinsic characteristic equation (3.10), or in other words
that the killing vector $k^\mu$ itself is bicharacteristic, in the sense of
being directed allong the propagation direction of extrinsic perturbations
of the world sheet. The criticality condition (6.37) is thus interpretable
as a condition of {\it characteristic flow}. It means that the ``running
velocity'', $v$ say, of relative motion of the intrinsicly preferred rest
frame of the string (as determined by whichever of $c^\mu$ and $\tilde
c^\mu$ is timelike) relative to the background frame specified by the (in
this case necessarily timelike) Killing vector $k^\mu$ is the same as the
extrinsic propagation velocity $c_{_{\rm E}}$ given by (4.10). 

In the presence of generic gravitational, electromagnetic and Kalb Ramond
forces, the criticality condition (6.37) can be satisfied at particular
positions, such as where there is a transition from a subcharacteristic
running velocity, $v<c_{_{\rm E}}$ to a supercharacteristic running velocity
$v>c_{_{\rm E}}$ (as will occur for instance on a string in a steady state
of radial flow into a Schwarzschild black hole) or where there is a cusp,
with $v=c_{_{\rm E}}$ but with subcharacteristic flow $v<c_{_{\rm E}}$ on
both sides. However in view of the constancy of the Bernoulli ``tuning"
parameters $\omega$ and $\beta$, the absence of any background field will
always allow, and generically (the special integrable case (5.49) being an
exception) will ensure, {\it uniformity} of the state of the string, so that
the transcharacteristic flow condition (6.37) can be satisfied throughout
its length. The space configuration of such a uniformly transcharacteristic
steady string state can have arbitrarily variable curvature, and so is
compatible with a closed loop topology. 

The question of closed loop equilibrium states did not arise in the earliest
studies\cite{[108]} of cosmic strings, which were restricted to the
Goto-Nambu model whose bicharacteristics are always null and so can never be
aligned with the timelike Killing vector generating a stationary symmetry.
However in a generic string model{[30]} for which the 2-dimensional
longitudinal Lorentz of the internal structure is broken by a current,
whether of the neutral kind exemplified in an ordinary violin string or the
electromagnetic kind exemplified\cite{[83]}\cite{[64]}\cite{[65]}\cite{[66]}
by Witten's superconducting cosmic string model, the bicharacteristic
directions will generically be timelike\cite{[36]} (spacelike
bicharacteristics being forbidden by the requirement of causality) so there
will be no obstacle to their alignement with a timelike Killing vector in
accordance with the criticality condition (6.37), i.e. to having a running
velocity given by $v= c_{_{\rm E}}$. The simple ``toy" complex scalar field
model on which the pionneering cosmic string studies\cite{[5]} were based
had longitudinally Lorentz invariant vortex defects (of ``local" or
``global" type depending on the presence or absence of coupling to a gauge
field) that were describable at a microscopic level by string models (with
Kalb Ramond coupling in the ``global" case) that were indeed of the special
Goto-Nambu type. However the extra degrees of field freedom (starting with
the additional scalar field introduced by Witten in his original
superconducting example\cite{[7]}) that are needed in successively more
realistic models\cite{[109]}\cite{[110]} make it increasingly difficult to
avoid the formation of internal structure breaking the longitudinal Lorentz
invariance and reducing the extrinsic characteristic velocity to the
subluminal range $c_{_{\rm E}}<1$ at which stationary equilibrium with the
critical running velocity $v=c_{_{\rm E}}<1$ becomes possible.

The cosmological significance of this is that whereas Goto Nambu string
loops cannot ultimately avoid gravitationally or otherwise radiating away
all their energy\cite{[111]}\cite{[112]}\cite{[113]}, since they have no
equilibrium states into which they might settle down, on the other hand more
general kinds of strings, whose occurrence would now seem at least as
plausible\cite{[109]}\cite{[110]}, can leave a relic distribution of
stationary loop configurations that may survive
indefinitely\cite{[105]}\cite{[114]}\cite{[115]}\cite{[116]}. One would
expect such configurations to be those that minimise the energy for given
values of the relevant globally conserved quantities, of which there might
be a considerable number in the more complicated multiply conducting models
that might be considered, but of which there are only a single pair in the
``barotropic" type string considered here, namely the stream function
winding number $\oint d\psi$ and the phase winding number $\oint d\varphi$
whose respective constancy results from the conservation of the mutually
dual pair of currents $c^\mu$ and $\tilde c^\mu$.

The potential cosmological importance of such a distribution of relic loops
was first pointed out by Davis and
Shellard\cite{[105]}\cite{[114]}\cite{[115]}, who emphasised that in the
case of the ``heavy" cosmic strings whose existence had been postulated to
account for galaxy formation, the ensuing relics, even if formed with very
low efficiency, would be more that sufficient to give rise to a catastrophic
cosmological mass excess of the kind first envisaged as arising from the
formation of monopoles. According to a more recent and detailed order of
magnitude estimate of my own\cite{[116]}, the dimensionless coupling
constant $Gm^2\approx 10^{-6}$ characterising ``heavyweight" strings,
meaning those arising from G.U.T. symmetry breaking (for which the relevant
Higgs mass scale $m$ is within a factor of order a thousand of the Plank
mass) would have to be reduced below a value given very roughly by
$Gm^2\approx 10^{-26}$ to avoid a cosmological mass excess today. This
estimate\cite{[116]} (based on considerations of the kind discussed during
the present meeting by Zurek) should be  regarded as provisional, pending
the more deeper investigation that would seem to be needed. What is
remarkable about this tentative limit is that it corresponds to a Higgs
mass scale $m$ of roughly the same order as that at which {\it electroweak}
symmetry breaking is believed to occur, which is rather suggestive in view
of the fact that although the ``standard" electroweak model does not give
rise to stable string like vortex defects\cite{[117]}, nevertheless such
defects do occur in many of its most commonly considered
competitors$\cite{[109]}$. 

The cosmological implications (as an argument against ``heavy" cosmic string
formation, or more positively as source of ``dark matter" in the form of
``lightweight" string loops due to electroweak symmetry breaking) of the
long term survival of cosmic string loops, motivates more thorough
investigation of equilibrium states that may be involved, a particularly
important question being that of their stability. Prior to the derivation of
the general symmetric string generator equation (6.26), the only closed loop
equilibrium states to have been considered were the circular kind referred
to by Davis and Shellard\cite{[105]}\cite{[114]}\cite{[115]} as ``vortons" a
term that is more appropriate for axisymmetric ring states in the ``global"
case than in the gauge coupled ``local" case for which the string
description is most accurate. The first general
investigation\cite{[4]}\cite{[118]} of such circular ring states showed that
under conditions of purely centrifugal support (neglecting possible
electromagnetic corrections of the kind evaluated more recently\cite{[107]})
the condition (6.37) for equilibrium, namely the requirement of a
transcharacteristic rotation speed $v=c_{_{\rm E}}$, is such that the ring
energy is minimised with respect to perturbations preserving the circular
symmetry. However a more recent investigation of non-axisymmetric
perturbations has shown that although there are no unstable modes for states
of subsonic rotation\cite{[67]} (as exemplified by a cowboy's lassoe loop)
with $v=c_{_{\rm E}}<c_{_{\rm L}}$, instability can nevertheless occur for
rotation in the supersonic regime\cite{[68]} that (contrary to what was
implicitly assumed in earlier 
work\cite{[52]}\cite{[53]}\cite{[54]}\cite{[55]}\cite{[56]}\cite{[57]}\cite{[58]} 
using the subsonic type of model given by a linear equation of state for
which the sum $U+T$ is constant) has been shown by
Peter\cite{[64]}\cite{[65]}\cite{[66]} to be relevant in the kind of cosmic
vortex defects that have been considered so far. Although the first category
of string loop equilibrium states to have been studied systematically has
been that of circular ring configurations\cite{[118]}, it has been made
clear by recent work\cite{[92]}\cite{[67]} that, as explained above,
arbitrary non circular equilibrium states are also possible. The stability
such more general equilibrium states has not yet been investigated,
carried out, but it seems plausible that while some of them may be 
destroyed by the
recently discovered classical instability mechanism\cite{[67]}\cite{[68]},
and also the kind of quantum tunnelling instability mechanism considered by
Davis\cite{[114]}, it does not seem likely that such mechanisms
could be so consistently efficient as to prevent the long term survival of
a lot of other stationary  loop states.

\vskip 1cm\noindent
{\bf Acknowledgements.}
\bigskip\noindent

I wish to thank B. Allen, C. Barrab\`es, U. Ben-Ya'acov, A-C. Davis, 
R. Davis, V. Frolov, G. Gibbons, R. Gregory, T. Kibble, K. Maeda, X. Martin, 
P. Peter, T. Piran, D. Polarski, M. Sakellariadou, P. Shellard, 
P. Townsend, N. Turok, T. Vachaspati, and  A. Vilenkin, for many
stimulating or clarifying discussions.

%\nonumsection
%{References}


\begin{thebibliography}{000}
 
\bibitem{[1]} P.A.M. Dirac, {\it Proc. Roy. Soc. London} {\bf A268}, 57 (1962).
%\smallskip
\bibitem{[2]} P.S. Howe, R.W. Tucker, {\it J. Phys.} {\bf A10}, L155 (1977).
%\smallskip
\bibitem{[3]} A. Ach\'ucarro, J. Evans,  P.K. Townsend, D.L. Wiltshire,
{\it Phys. Lett.} {\bf 198 B}, 441 (1987).
%\smallskip
\bibitem{[4]}  B. Carter,   in {\it Formation and Evolution of Cosmic Strings,} 
ed. G. Gibbons, S. Hawking, T. Vachaspati, pp143-178 (Cambridge U.P., 1990).
%\smallskip
\bibitem{[5]} T.W.B. Kibble, {\it J.Phys.} {\bf A9}, 1387 (1976).
%\smallskip
\bibitem{[6]} A. Vilenkin, A.E. Everett, {\it Phys. Rev. Lett} {\bf 48}, 1867 
(1982).
%\smallskip
\bibitem{[7]} E. Witten, {\it Nucl. Phys.} {\bf B249}, 557 (1985).
%\smallskip
\bibitem{[8]} B. Carter, {\it J. Geom. Phys.} {\bf 8}, 53 (1992).
%\smallskip
\bibitem{[9]} B. Carter, {\it J. Class. Quantum Grav.} {\bf 9}, 19 (1992).
%\smallskip
\bibitem{[10]} L.P. Eisenhart, {\it Riemannian Geometry} (Princeton U.P., 
1926, reprinted 1960).
%\smallskip
\bibitem{[119]} J. Stachel, {Phys. Rev.} {\bf D21}, 2171 (1980).
%\smallskip
\bibitem{[11]} J.A. Schouten, {\it Ricci Calculus} (Springer, Heidelberg, 1954).
%\smallskip
\bibitem{[12]} B. Carter, M. Sakellariadou, X. Martin, {\it Phys. Rev.} 
{\bf D50}, 682 (1994)
%\smallskip
\bibitem{[13]} A. Vilenkin, {\it Phys. Rev.} {\bf D43}, 1060 (1991).
%\smallskip
\bibitem{[14]} R. Penrose, W. Rindler, {\it Spinors and Space-Time} (Cambridge
U.P., 1984).
%\smallskip
\bibitem{[15]} I. Bars, C.N. Pope, {\it Class. Quantum Grav.,} {\bf 5}, 1157 
(1988).
%\smallskip
\bibitem{[16]} P. Sikivie, {\it Phys. Rev. Lett.} {\bf 48}, 1156 (1982).
%\smallskip
\bibitem{[17]} A. Vilenkin, A.E. Everett,{\it Phys. Rev. Lett.} {\bf 48}, 1867
(1982).
%\smallskip
\bibitem{[18]}  E.P.S. Shellard,   in {\it Formation and Evolution of Cosmic
Strings,} ed. G. Gibbons, S. Hawking, T. Vachaspati, pp107-115 
(Cambridge U.P., 1990).
%\smallskip
\bibitem{[19]} Y. Nambu, {\it Nucl. Phys.} {\bf B130}, 505 (1977).
%\smallskip
\bibitem{[20]} N.S. Manton {\it Phys. Rev.} {\bf D28}, 2019 (1983).
%\smallskip
\bibitem{[21]} T. Vachaspati, A. Ach\'ucarro, {\it Phys. Rev.} {\bf D44}, 3067
(1991).
%\smallskip 
\bibitem{[22]} T. Vachaspati, M. Barriola, {\it Phys. Rev. Letters} {\bf 69},
1867 (1992).
%\smallskip
\bibitem{[23]}  A. Dabholkar, J.M. Quashnock, {\it Nucl. Phys.} {\it B333},
 815 (1990).
%\smallskip
\bibitem{[24]} R.A. Battye, E.P.S. Shellard, {\it 
``Global string radiation"}, DAMTP preprint (Cambridge, 1994).  
%\smallskip
\bibitem{[25]} B. Carter, {\it Class. Quantum Grav.} {\bf 11}, 2013 (1994).
%\smallskip
\bibitem{[26]} A. Vilenkin, T. Vachaspati, {\it Phys. Rev.} {\bf D35}, 1138 
(1987).
%\smallskip
\bibitem{[27]} R.L. Davis, E.P.S. Shellard, {\it Phys. Rev. Lett.} {\bf 63}, 
2029 (1989).
%\smallskip
\bibitem{[28]} M. Sakellariadou, {\it Phys. Rev.} {\bf D44}, 3767 (1991).
%\smallskip
\bibitem{[29]} U. Ben-Ya'acov, {\it Nucl. Phys.} {\bf B382}, 597 (1992).                                        
%\smallskip
\bibitem{[30]} B. Carter, {\it Phys. Lett.} {\bf B224}, 61 (1989).
%\smallskip
\bibitem{[31]} A.L. Larsen, {\it Class. Quantum Grav.} {\bf 10L}, 35 (1993).
%\smallskip
\bibitem{[32]} J. Garriga, M. Sakellariadou, {\it Phys. Rev.} {\bf 48}, 2502
 (1993).
%\smallskip
\bibitem{[36]} B. Carter, {\it Phys. Lett.} {\bf B228}, 466 (1989).
%\smallskip
\bibitem{[33]} B. Carter, {\it Phys. Rev.} {\bf 48}, 4835 (1993)
%\smallskip
\bibitem{[59]} J. Garriga, A. Vilenkin, {\it Phys. Rev.} {\bf D47},
3265 (1993).
%\smallskip
\bibitem{[34]} A.L. Larsen, V. Frolov, {\it Nucl. Phys.} {\bf B414}, 129
(1994).
%\smallskip
\bibitem{[60]} J. Garriga, A. Vilenkin, {\it Phys. Rev.} {\bf D44}, 1007
(1991).
%\smallskip
\bibitem{[61]} J. Garriga, A. Vilenkin, {\it Phys. Rev.} {\bf D45}, 3469
(1992).
%\smallskip
\bibitem{[62]} J. Guven, {\it Phys. Rev.} {\bf D48}, 4464 (1993).
%\smallskip
\bibitem{[63]} J. Guven, {\it Phys. Rev.} {\bf D48}, 5563 (1993).
%\smallskip
\bibitem{[35]} B. Carter, {\it Proc. Roy. Soc. Lond.} {\bf A372},
169 (1980).
%\smallskip
\bibitem{[37]} A. Polyakov, {\it Nucl. Phys.} {\bf B268}, 406 (1986).
%\smallskip
\bibitem{[38]} K.I. Maeda, N. Turok, {\it Phys. Lett.} {\bf B202}, 376 
(1988). 
%\smallskip
\bibitem{[39]} R. Gregory, {\it Phys. Lett.} {\bf B206}, 199 (1988). 
%\smallskip
\bibitem{[40]} R. Gregory, {\it Phys. Rev.} {\bf D43}, 520 (1993).
%\smallskip
 \bibitem{[41]} D. Garfinkle, R. Gregory, {\it Phys. Rev.} {\bf D41}
 1889 (1990).
%\smqllskip
\bibitem{[42]} R. Gregory, D. Haws, D. Garfinkle, {\it Phys. Rev.} {\bf D42}
343 (1991).
%\smallskip
\bibitem{[43]} V. Silveira, M.D. Maia, {\it Phys. Lett} {\bf A174}, 280 
(1993).
\bibitem{[44]} C. Barrab\`es, B. Boisseau, M. Sakellariadou, {\it Phys. Rev.}
{\bf D49}, 2734 (1994).
%\smallskip
\bibitem{[45]} B. Carter, R. Gregory, {\it Curvature corrections to 
dynamics of domain walls}, DAMTP preprint (Cambridge, 1994).
%\smallskip
\bibitem{[46]} P.S. Letelier, {\it Phys. Rev.} {\bf D41}, 1333 (1990).
%\smallskip
\bibitem{[47]} D. H. Hartley, R.W. Tucker, in {\it Geometry of Low Dimensional
Manifolds, 1 }( {L.M.S. Lecture Note Series} {\bf 150}, 
ed. S. Donaldson, C. Thomas (Cambridge U.P., 1990).
%\smallskip
\bibitem{[120]}  H. Arodz, A. Sitarz, P. Wegrzyn, {\it Act. Phys.
Polon. B}, {\bf 22}, 495 (1991); {\bf 23}, 53 (1992). 
%\smallskip
\bibitem {[48]} B. Boisseau, P.S. Letelier, {\it Phys. Rev.} {\bf D46}, 1721
(1992).
%\smallskip
\bibitem {[49]} B. Carter, {\it Equations of motion of a stiff geodynamic
 string or higher brane}, Meudon preprint (to appear in {\it Class. 
Quantum Gravity} {\bf 11}, 1994).
%\smallskip
\bibitem{[50]} S.W. Hawking, G.F.R. Ellis, {\it The Large Scale Structure of
Spacetime} (Cambridge U.P., 1973).
%\smallskip
\bibitem{[51]} B. Carter, in {\it Active Galactic Nuclei}, eds.
C. Hazard, S.Mitton, pp 273-300 (Cambridge U. P., 1976).
%\smallskip
\bibitem{[52]} D.N. Spergel, T. Piran, J. Goodman, {\it Nucl. Phys.}
{\bf B291}, 847 (1987).
%\smallskip
\bibitem{[53]} A. Vilenkin, T. Vachaspati, {\it Phys. Rev. Lett.} {\bf 58},
1041 (1987)
%\smallskip
\bibitem{[54]} E. Copeland, M. Hindmarsh, N. Turok, {\it Phys. Rev. Lett.}
{\bf 58}, 1910 (1987)
%\smallskip
\bibitem{[55]} E. Copeland, D. Haws, M. Hindmarsh, N. Turok, {\it Nucl.
Phys.} {\bf B306}, 908 (1988).
%\smallskip
\bibitem{[56]} D. Haws, M. Hindmarsh, N. Turok, {\it Phys. Lett.} 
{\bf B209}, 225 (1988).
%\smallskip
\bibitem{[57]} D.N. Spergel, W.H. Press, R.J. Scherrer, {\it Phys. Rev.}
{\bf D39}, 379 (1989)
%\smallskip
\bibitem{[58]} P. Amsterdamski, {\it Phys Rev.} {\bf D39}, 1534 (1989).
%\smallskip 
\bibitem{[64]} P. Peter, {\it Phys. Rev.} {\bf D45}, 1091 (1992).
%\smallskip
\bibitem{[65]} P. Peter, {\it Phys. Rev.} {\bf D46}, 3335 (1992).
%\smallskip
\bibitem{[66]} P. Peter, {\it Phys. Rev.} {\bf D47}, 3169 (1993).
%\smallskip
\bibitem{[67]} B. Carter, X. Martin, {\it Ann. Phys.} {\bf 227}, 151
(1993).
%\smallskip
\bibitem{[68]} X. Martin, Meudon preprint (to appear in {\it Phys. Rev.}
{\bf D}, 1994).
%\smallskip
\bibitem{[69]} R.L. Davis, E.P.S. Shellard, {\it Phys. Lett} {\bf B209},
485 (1988).
%\smallskip
\bibitem{[70]} R.L. Davis, E.P.S. Shellard, {\it Nucl. Phys.} {\bf B323}
209 (1989)
%\smallskip
\bibitem{[71]} B. Carter, {\it Phys. Lett.} {\bf B238}, 166 (1990). 
%\smallskip
\bibitem{[72]} P. Peter, {\it Phys. Lett.} {\bf B298}, 60 (1993). 
%\smallskip
\bibitem{[73]} N.K. Nielsen, {\it Nucl. Phys.} {\bf B167}, 248 (1980).
%\smallskip
\bibitem{[74]} N.K. Nielsen, P. Olesen, {\it Nucl. Phys.} {\bf B291},
829 (1987). 
%\smallskip
\bibitem{[75]} A. Davidson, A.K. Wali, {\it Phys. Lett.} {\bf B213}, 419 
(1988).
%\smallskip
\bibitem{[76]} A. Davidson, A.K. Wali, {\it Phys. Rev. Lett.} {\bf 61},
1450 (1988).
%\smallskip
\bibitem{[77]} B. Carter, {\it Phys. Rev.} {\bf D41}, 3869 (1990).
%\smallskip
\bibitem{[78]} A. Vilenkin, {\it Phys. Rev.} {\bf D41}, 3038 (1990).
%\smallskip
\bibitem{[79]} J.Hong, J.Kim, P. Sikivie, {\it Phys. Rev. Lett.} {\bf 69},
2611 (1980).
%\smallskip
\bibitem{[80]} B. Carter, {\it Nucl. Phys.} {\bf B412}, 345 (1994).
%\smallskip
\bibitem{[81]} E. Witten, {\it Phys. Lett.} {\bf B158}, 243 (1985).
%\smallskip
\bibitem{[82]} R.L. Davis, E.P.S. Shellard, {\it Phys. Lett.} {\bf B214},
219 (1988).
%\smallskip
\bibitem{[83]} A. Babul, T. Piran, D.N. Spergel, {\it Phys. Lett.}
{\bf B202}, 207 (1988).
%\smallskip
\bibitem{[84]} D.N. Spergel, W.H. Press, R.J. Scherrer, {\it Phys. Rev.}
{\bf D39}, 379 (1989).
%\smallskip
\bibitem{[85]} T.W.B. Kibble, N. Turok, {\it Phys. Lett.} {\bf B116},
141 (1982).
%\smallskip
\bibitem{[86]} X. Martin, Meudon preprint (1994).
%\smallskip
\bibitem{[87]} B. Allen, E.P.S. Shellard, {\it Phys. Rev. Lett.}
{\bf 64}, 119 (1990).
%\smallskip
\bibitem{[88]}  E.P.S. Shellard, B. Allen, in {\it Formation and Evolution
of Cosmic Strings,} ed. G. Gibbons, S. Hawking, T. Vachaspati, pp421-448
(Cambridge U.P., 1990).
%\smallskip
\bibitem{[89]} A.G. Smith, A. Vilenkin, {\it Phys. Rev.} {\bf D36}, 990
(1987).
%\smallskip
\bibitem{[90]}  A. Albrecht, in {\it Formation and Evolution
of Cosmic Strings,} ed. G. Gibbons, S. Hawking, T. Vachaspati, pp403-419
(Cambridge U.P., 1990).
%\smallskip
\bibitem{[91]} B. Carter, {\it Class. and Quantum Grav.} {\bf 7}, L69
(1990).
%\smallskip
\bibitem{[92]} B. Carter, V.P. Frolov, O. Heinrich, {\it Class. and Quantum 
Grav.}  {\bf 8}, 135 (1991).
%\smallskip

\bibitem{[93]} V.P. Frolov, V.D. Skarzhinsy, A.I. Zelnikov, O. Heinrich,
{\it Phys. Lett.} {\bf 224}, 225 (1989).
%\smallskip
\bibitem{[94]} B. Carter, V.P. Frolov, {\it Class. Quantum Grav.} {\bf 6},
569 (1989).
%\smallskip
\bibitem{[95]} B. Carter, in {\it Black Holes (Les Houches, 1972)}, 
ed. C. and B.S. DeWitt, pp 57-214 (Gordon and Breach, New York, 1973).
%\smallskip
\bibitem{[96]} G.W. Gibbons, S.W. Hawking, {\it Phys. Rev.} {\bf D15}, 2738
(1976).
%\smallskip
\bibitem{[97]} B. Carter, {\it Phys. Rev.} {\bf 174}, 1559 (1968).
%\smallskip
\bibitem{[98]} B. Carter, {\it Commun. Math. Phys.} {\bf 99},  563 (1968).
%\smallskip 
\bibitem{[99]} S.A. Teukolsky, {\it Astroph. J.} {\bf 185}, 283 (1973).
%\smallskip
\bibitem{[100]} W. Unruh, {\it Phys. Rev. Lett.} {\bf 31}, 1265 (1973).
%\smallskip
\bibitem{[101]} S. Chandrasekhar, {\it Proc. Roy. Soc. Lond.} {\bf A349},
571 (1976).
%\smallskip
\bibitem{[102]} R. Guven, {\it Phys. Rev.} {\it D22}, 2327 (1980).
%\smallskip
\bibitem{[103]} J-A. Marck, {\it Proc. Roy. Soc. Lond.} {\bf A385}, 431
(1983).
%\smallskip
\bibitem{[104]} B. Carter, {\it J. Math. Phys.} {\bf 28}, 1535 (1987).
%\smallskip 
\bibitem{[105]} R.L. Davis, E.P.S. Shellard, {\it Phys. Lett.}
{\bf B209}, 485 (1988).
%\smallskip
\bibitem{[106]} J. Ostriker, C. Thompson, E. Witten, {\it Phys. Lett} 
{\bf B180}, 231 (1986)
%\smallskip
\bibitem{[107]} P. Peter, {\it Phys. Lett.} {\bf B298}, 60 (1993).
%\smallskip
\bibitem{[108]} T.W.B. Kibble, {\it Phys. Rep.} {\bf 67}, 183 (1980).
%\smallskip
\bibitem{[109]} P. Peter,  {\it Phys. Rev.}, {\bf D46}, 3322 (1992).
%\smallskip
\bibitem{[110]} W. B. Perkins, A-C. Davis, {\it Nucl. Phys.} {\bf B406},
377  (1993).
%\smallskip
\bibitem{[111]} N. Turok, {\it Nucl. Phys.} {\bf B242}, 520 (1984).
%\smallskip
\bibitem{[112]} T. Vachaspati, A. Vilenkin, {\it Phys Rev.} {\bf D31}, 
3035, (1985).
%\smallskip
\bibitem{[113]} R. Durrer, {\it Nucl. Phys.} {\bf B328}, 238 (1989).
%\smallskip
\bibitem{[114]} R.L. Davis, {\it Phys. Rev.} {\bf D38}, 3722 (1988).
%\smallskip
\bibitem{[115]} R.L. Davis, E.P.S. Shellard, {\it Nucl. Phys.} {\bf B323},
209 (1989).
%\smallskip
\bibitem{[116]} B. Carter, {\it Ann. N.Y. Acad. Sci.} {\bf 647}, 758 
(1991).
\bibitem{[117]} M. James, L. Perivolaropoulos, T. Vachaspati, {\it Nucl.
Phys.} {\bf B395}, 534 (1993).
%\smallskip
%\smallskip
\bibitem{[118]} B. Carter, {\it Phys. Lett.} {\bf B238}, 166 (1990).
%\smallskip
                                                        
\end{thebibliography}
 \end{document}